\documentclass[11pt,letterpaper]{article}
\pdfoutput=1
\usepackage{jheppub}

\usepackage[english]{babel}
\usepackage{amssymb}
\usepackage{amsfonts}
\usepackage{amsmath}
\renewcommand\d{\partial}

\newcommand{\be}{\begin{equation}} \newcommand{\ee}{\end{equation}}
\newcommand{\bea}{\begin{eqnarray}} \newcommand{\eea}{\end{eqnarray}}
\newcommand{\beann}{\begin{eqnarray*}}  \newcommand{\eeann}{\end{eqnarray*}}
\newcommand{\bfig}{\begin{figure}} \newcommand{\efig}{\end{figure}}
\newcommand{\ba}{\begin{array}} \newcommand{\ea}{\end{array}}
\newcommand{\bcen}{\begin{center}} \newcommand{\ecen}{\end{center}}
\newcommand{\btab}{\begin{tabular}} \newcommand{\etab}{\end{tabular}}
\newcommand{\nn}{\nonumber}
\title{Non-relativistic holography from Ho\v rava gravity}

\author{Stefan Janiszewski}
\author{and Andreas Karch}

\affiliation{Department of Physics, University of Washington, Seattle, WA
98195-1560, USA}

\emailAdd{stefanjj@u.washington.edu}
\emailAdd{akarch@u.washington.edu}

\abstract{
Many non-relativistic Quantum Field Theories with conserved particle number share a common set of symmetries: time dependent spatial diffeomorphisms acting on the background metric and $U(1)$ invariance acting on the background fields which couple to particle number. Here we use these symmetries to deduce a gravity dual for any such theory in terms of a non-relativistic theory of gravity, a variant of Ho\v rava gravity. This duality allows the extension of holography to generic non-relativistic field theories. As Ho\v rava gravity is presumed to be a consistent quantum theory, this duality also allows holography to move beyond the limit of a large number of colors, in principle. In the case when the field theory is conformally invariant, we prove that our proposal reproduces the form of the two point function demanded by this symmetry.
}

\begin{document}
\maketitle
\flushbottom

\section{Introduction}
\subsection{Non-relativistic quantum field theories}\label{sec:NRQFTintro}

While diffeomorphism invariance, the statement that physics does not depend on the coordinate system used for spacetime, is often seen as a hallmark of the theory of general relativity, it is already a property of any relativistic quantum field theory (QFT) formulated on a fixed, potentially curved, background spacetime metric. In the case of a QFT, diffeomorphism invariance should be seen as a ``global symmetry,'' in contrast to a gauge symmetry. The latter is not a symmetry at all, but a redundancy in the description. Gauge variant quantities are simply not physical. The gauge variant description introduced non-physical degrees of freedom to simplify the Lagrangian; the gauge invariance of observables removes those extra degrees of freedom again. Diffeomorphisms are a gauge invariance of gravity. This is the reason that gravity has no standard local observables.

A global symmetry, on the other hand, is a real symmetry of the system. Physical quantities need to furnish a representation of it. Sometimes parameters of the theory transform non-trivially under global symmetries. This is, for example, the case in the theory of a massive Dirac fermion. The mass term explicitly breaks the axial symmetry. However, the symmetry can be restored if we assign the mass term axial charge. Although such ``spurionic'' symmetries do not in general generate conserved charges, they are useful as they constrain how couplings can appear in the low energy effective theory or correlation functions. In a QFT on a fixed spacetime background the metric acts analogously to the Dirac fermion mass. One should think of the metric as a set of coupling constants specified at every point in spacetime. Position dependent diffeomorphisms are now a global symmetry under which these coupling constants transform. Similarly, a background gauge field should be viewed as a set of coupling constants in the
Hamiltonian that transform non-trivially under a ``global" position dependent $U(1)$ transformation.

In a non-relativistic (NR) quantum field theory, time plays a special role: there is a preferred notion of spatial slices consisting of events happening simultaneously. This can be implemented by considering the spacetime manifold to be equipped with a co-dimension one foliation consisting of the spatial leaves. A global time defines the invariant notion of whether one event occurs before or after another, and is hence required in order to have a well defined causality. Non-relativistic theories can have instantaneous interactions that, when turned on, have immediate influence at arbitrarily large spatial distances, but they cannot influence events that occurred at an earlier global time. In this way causality is preserved in the absence of light cones.

Usually one wants to insist on translation invariance in time, $t \rightarrow \tilde{t} =t - f$ where $f$ is a constant, so that the system allows for a conserved energy. Sometimes one can extend this symmetry to include the case where $f$ is linear in $t$, or even to the case where $f$ is an arbitrary function of $t$. As we will review, these two special cases correspond to NR QFTs which are scale and conformally invariant, respectively. Any such time coordinate has the right to be called a global time: the leaves of the foliation remain at constant time even after the transformation $f(t)$. On the other hand, the Lorentzian diffeomorphism where $f$ has spatial dependence, violates the preferred foliation as it changes the time ordering of events. Such a redefined temporal coordinate cannot be considered a global time because it would alter the notion of which events occur before or after another, and hence violate causality.  Although in a NR QFT one can always work in a global time, and restrict $f$
to be a function of time only, insight and information can be gained by considering a ``non-physical'' time and allowing spatial dependence of $f$. This is analogous to using an arbitrary metric in a relativistic field's action so that one can calculate the stress-energy tensor, even if only interested in flat Minkowski space. As we will discuss, from the non-relativistic viewpoint one can still consider these non-physical temporal transformations by having them act on a background source coupling to energy current.

For a NR QFT defined in $d$ spatial dimensions, we however should still expect invariance under purely spatial diffeomorphisms. Furthermore, for many NR QFTs we are allowed to perform a different change of spatial coordinates at different times, that is $x_i \rightarrow \tilde{x}_i(x_i,t)$. In particular, these time dependent spatial diffeomorphisms include the standard Galilean boosts. Together with translations and rotations these boosts play a special role as they leave a flat space background with no electromagnetic field invariant. Correspondingly, they do not just constrain the low energy effective action but are true symmetries and give rise to conserved charges.

Additionally, most NR QFTs allow for a conserved particle number current. In this case we can also formulate the theory in the presence of background electric and magnetic fields coupling to particle number, and the theory possesses a position dependent $U(1)$ global symmetry acting on the associated background potential. We can take these symmetries --- time-dependent spatial diffeomorphisms, a $U(1)$ rotation acting on the background gauge field coupled to particle number, and time translation invariance --- as the defining symmetries of a large class of NR QFTs. This class includes most interacting electron systems and in particular the quantum Hall states.
If the theory, in addition, allows for arbitrary reparametrizations of time, it describes a conformal NR QFT, of which the unitary Fermi gas is an example.
For these conformal theories there again exists a subgroup of transformations that leaves the trivial field theory metric and gauge potential invariant. This subgroup is often referred to as the Schr\"odinger group.

These spurionic symmetries put strong constraints on possible terms in the low energy effective action in an interacting NR QFT. This was exploited for the unitary Fermi gas in \cite{Son:2005rv}, which also developed most of the formalism used here, and, more recently, for quantum Hall states in
\cite{Hoyos:2011ez}. For the quantum Hall states these symmetries allow one to relate the Hall viscosity and the change in filling fraction when the theory is put on a sphere to a single coefficient in the low energy effective action. Furthermore, the leading correction to the Hall conductivity in the presence of a background electric field with slow spatial variation is completely determined by the symmetry in terms of thermodynamic quantities. As the Hall states describe gapped states, the only fields appearing in the low energy effective action are the background metric and background electric fields, making symmetries very powerful. In the unitary Fermi gas the interplay between NR conformal invariance and NR diffeomorphisms constrains several transport coefficients in the hydrodynamic description of this system \cite{Son:2005rv}.

One can obtain a NR QFT by taking the speed of light $c\to\infty$ limit of a relativistic field theory. In order to yield non-trivial results, a chemical potential $\mu$ must be turned on to provide the rest mass $m$ of particles. This causes the free energy associated with a particle to remain finite in the large $c$ limit, while the free energy associated with an antiparticle goes to infinity as twice its rest energy and they therefore completely decouple. The absence of antiparticles in a NR QFT means that virtual pairs cannot lead to particle creation. Instead the existence of particles requires a chemical potential to pay their rest mass. The non-relativistic theory then describes fluctuations around this energy. We will make extensive use of this concept and the $c\to\infty$ limit.

\subsection{Holography}

Gauge-gravity duality \cite{Maldacena:1997re,Witten:1998qj,Gubser:1998bc}, or ``holography" for short, is a powerful tool that allows one to solve certain strongly coupled gauge theories in terms of a dual gravitational description in one higher dimension. For the gravitational theory to be classical it needs to have a large separation between the scale of curvature of the geometry and the Planck mass. In the field theory this requires a ``large $N_c$" limit, where $N_c$ is the number of colors if the field theory is a non-Abelian gauge theory, or a similar measure of the number of degrees of freedom in other cases. One very basic piece of evidence for this equivalence is the matching of symmetries on both sides. For relativistic field theories formulated on a fixed background geometry this includes changes of coordinates on this background spacetime metric. In the bulk these diffeomorphisms are part of
the higher dimensional diffeomorphism invariance. Bulk diffeomorphisms that vanish near the boundary of the space correspond to a gauge invariance in the bulk. As usual, they should not be interpreted as a global symmetry. They correspond to a redundancy in the description of the bulk theory. However, changes of coordinates that do not depend on the extra holographic radial direction do correspond to global symmetries as they act on the boundary data in the expected way: they correspond to a diffeomorphism acting on the metric the field theory lives on.

As most strongly coupled systems of interest in condensed matter physics are non-relativistic, there has recently been much interest in formulating holography for NR QFTs, starting with the work of \cite{Son:2008ye,Balasubramanian:2008dm}. The theories studied in those works enjoyed the full symmetry of a NR conformal field theory (CFT). More importantly, these NR CFTs were obtained by a light-like reduction of a relativistic CFT, from which they inherited most of their properties. They hardly constitute generic NR QFTs.

As in the relativistic case, we believe a guiding principle for constructing a gravitational dual should be the defining symmetries of a generic NR QFT. A holographic gravity dual should have the same set of symmetry transformations as the field theory we are interested in: time dependent spatial diffeomorphisms, spatially dependent temporal diffeomorphisms, and the $U(1)$ symmetry acting on the background gauge field coupled to particle number. We will refer to this set of transformations  as ``NR electro-diffeomorphisms''. If we restrict the temporal diffeomorphisms, excluding the non-physical spatially dependent ones that violate the preferred foliation, we have the ``NR general covariance'' of \cite{Horava:2010zj}. Furthermore, if we exclude the $U(1)$ gauge symmetry we have the ``foliation preserving diffeomorphisms'' of \cite{Horava:2009uw}. We emphasize that any NR QFT that has NR electro-diffeomorphisms as its symmetry group must still have a notion of global time in order to have a well
defined causality. This means the spacetime manifold comes equipped with a foliation by spatial leaves parametrized by a global time. Such a theory can
therefore be restricted to have only NR general covariance by working in coordinates adapted to the foliation. Although the symmetry group of NR electro-diffeomorphisms can give us more information about a theory, it can only describe the same causal theories that NR general covariance can.

A gravitational theory centered around foliation preserving diffeomorphisms was introduced by Ho\v rava in \cite{Horava:2009uw}. In its most simple form, Ho\v rava-Lifshitz theory describes the dynamics of a lapse field $N$, a shift vector $N_{I}(t,x_I)$, and a spatial metric\footnote{We are using indices $i$, $j$, $\ldots$ running over the $d$ spatial dimensions of the field theory; $\mu$, $\nu$, $\ldots$ running over the $d+1$ field theory directions including time; $I$, $J$, $\ldots$ running over the $d$+1 spatial dimensions of the bulk including the radial coordinate $r$; and, last but not least, $M$, $N$, $\ldots$ running over all $d+2$ bulk directions including time and $r$. In section \ref{sec:kkk} we will require discussion of a $d+3$ dimensional bulk, there we use indices $X$, $Y$, $\ldots$ to cover the $d+2$ directions of $M$, $N$, $\ldots$ plus one additional direction $\zeta$.} $G_{IJ}(t,x_I)$. In the language of \cite{Horava:2009uw} the theory is ``projectable" if $N$ is a
function of $t$ only, and non-projectable when $N$ is allowed to have spatial dependence as well. Writing the most general low energy action consistent with symmetries and containing up to two derivatives, one
finds that the action is almost completely fixed to be that of Einstein gravity written in terms of these fields. In addition to the two free dimensionful parameters of Einstein's gravity, the Newton's constant $G_N$ and the cosmological constant $\Lambda$, the low energy limit of projectable Ho\v rava gravity has one additional free parameter $\tilde{\lambda}$, which determines the relative coefficient of the two allowed kinetic terms for the spatial metric, written in terms of the extrinsic curvature of the slice. In the non-projectable case, which will be the main interest in this work, there is another  two derivative term we can include in the low energy action involving spatial derivatives of $N$. The corresponding coupling constant is commonly referred to as $\alpha$. These parameters are one of the issues that makes it difficult to find a version of Ho\v rava gravity that is a consistent theory of our world. In order to agree with the observed Lorentz invariance one needs a mechanism to set $\tilde{\lambda}\approx \alpha \approx 0$, the
value they take in general relativity (GR). For applications to NR holography, this is of no concern. In fact, one could hope that by adjusting these couplings Ho\v rava gravity could holographically describe a wide class of NR QFTs.

The projectable version of Ho\v rava-Lifshitz theory was extended in \cite{Horava:2010zj,daSilva:2010bm} to include NR general covariance (that is, the $U(1)$ symmetry corresponding to particle number conservation in addition to foliation preserving diffeomorphisms). In this case, the theory contains two additional non-dynamical fields, the ``potential" $A(t,x_I)$ (which arises as the subleading term of $N$ in a non-relativistic expansion and in that sense it restores spatial dependence to $N$) and at least one of the following: a field $A_{IJ}$, which can be thought of as the subleading term of the spatial metric, or the so called ``prepotential" field $\nu(t,x_I)$. The one exception is the case of $D=2$ spatial dimensions, for which no extra field beyond $A$ is required. These versions do not allow a straightforward holographic interpretation. In $D=2$ dimensions the equations of motion for $A$ immediately force spatial slices to be flat, whereas for holographic
interpretations following the standard recipe we expect an asymptotically hyperbolic spatial slice\footnote{AdS in flat slicing has been found as a solution to projectable Ho\v rava gravity \cite{Greenwald:2010fp}, but given in the Gullstrand-Painleve coordinates, which do not extend to the boundary. These are related to the traditional Fefferman-Graham coordinates \cite{Fefferman:1985} by a ``non-physical'' temporal transformation, and so correspond to gauge inequivalent configurations of Ho\v rava gravity.}. Similarly, the theory with $A_{IJ}$ requires a flat spatial slice\footnote{While it is possible to introduce a ``spatial cosmological constant" $\Omega$ in the theory with $A_{IJ}$, the constraints that arise as the equations of motion of $A$ and $A_{IJ}$ are only satisfied if $\Omega=0$.}.
The scenario with the prepotential $\nu$ has a different problem. Under the $U(1)$ symmetry $\nu$ shifts. Therefore, as discussed more in section \ref{sec:scalar}, the $U(1)$ gauge invariance in the bulk is completely fixed by choosing $\nu=0$ gauge; there are
no residual gauge transformations left that could be interpreted as global symmetries acting on the background data of the dual field theory. One could instead adopt the $N_r=0$ gauge, which leaves $r$-independent gauge transformations as a residual symmetry. In this case the asymptotic value $\nu$ would have to be interpreted as the source of a boundary operator. Like the background electric and magnetic fields, this background coupling constant would not be invariant under the $U(1)$ global transformation. Unlike the former, which do transform exactly like background fields should under a $U(1)$ transformation, $\nu$ shifts also in the boundary theory. The only example of an operator that transforms like this, that we are aware of, would be the phase of a $U(1)$ charged operator; if either added to the Lagrangian or having acquired an expectation value the presence of this operator would signal that in the boundary theory the $U(1)$ symmetry is broken (explicitly or spontaneously, respectively).
Thus the theory with $\nu$ can at best capture the dual to a NR QFT with a broken $U(1)$.

We will derive a different field content that obeys the symmetries of NR electro-diffeomorphisms by taking a particular Kaluza-Klein compactification of GR, as well as by taking the infinite speed of light limit of Einstein-Maxwell theory. The main thrust of this paper is that by working in adapted coordinates, and restricting the symmetry transformations to exclude the non-physical temporal diffeomorphisms, non-projectable Ho\v rava gravity coupled to electric and magnetic fields captures NR general covariance, and therefore should be dual to a generic NR QFT with these same symmetries.

Ho\v rava-Lifshitz theory comes with an intrinsic scale, the Planck mass $M_{pl}$. For energies far below the Planck mass the action should be limited to 2-derivative terms and is uniquely fixed (given $G_N$ and $\Lambda$) up to a small set of free parameters: $\tilde{\lambda}$ and $\alpha$ introduced above. This is the appropriate action to use when $L$, the typical curvature radius of spacetime, is large in Planck units. From experience with relativistic holography, this limit corresponds to a large $N_c$ limit in the dual QFT, which allows one to study a classical bulk theory. One of the big selling points of Ho\v rava-Lifshitz theory is that it is a candidate for a UV finite quantum theory of gravity. At energies far above the Planck scale the theory is argued to flow to a UV fixed point with a different dynamical critical exponent\footnote{We use the notation $z_{HL}$ here for the dynamical critical exponent of the Ho\v rava gravity Lagrangian to distinguish it from the dynamical critical exponent $z$
of the
dual field theory which, as we will see, can vary even in the case of $z_{HL}=1$} $z_{HL}$. As a consequence, at this putative UV fixed point the counting of derivatives needs to distinguish between spatial derivatives, which have dimension 1, and temporal derivatives, which have dimension $z_{HL}$. All marginal and relevant terms (that is terms with dimension less than or equal to $D+z_{HL}$, which compensates the dimension $-D-z_{HL}$ of the integration measure $d^Dx \, dt$) need to be included in the action. In particular, the potential energy, which depends on the curvature of the spatial metric $G_{IJ}$ and its spatial derivatives, should include terms with up to $D+z_{HL}$ derivatives of the metric. For the special case of $D=3$, $z_{HL}=3$ a full list of the possible terms in the potential, subject to certain discrete symmetry assumptions, has been worked out\footnote{In the original work of \cite{Horava:2009uw} a simpler potential has been used by imposing the additional constraint of detailed
balance. It seems to still be under debate whether this constraint can be imposed at the full quantum level. This question is not relevant for the $M_{pl} L \gg 1$ case.} in \cite{Sotiriou:2009gy}. In this work we will instead focus on the low energy limit as appropriate when $M_{pl} L \gg 1$, that is, when the dual field theory is taken in the large $N_c$ limit. In this case, we are only interested in energies $E \sim 1/L \ll M_{pl}$  and only terms with up to two spatial derivatives can occur in the potential. Studying the low energy, large $N_c$ limit will allow us to firmly establish the dictionary between bulk and boundary quantities. The prospect that covariant Ho\v rava-Lifshitz theory at a given $z_{HL}$ may be a complete quantum theory, and hence allow us to study the dual NR QFT at finite $N_c$, makes this approach extremely promising and is something left for future exploration.

In the large landscape of internally consistent quantum field theories, the highly constrained class of relativistic quantum field theories occupies only a small corner. This work suggests that something similar should be true on the dual holographic side. While the well studied case of gravitational theories with the full relativistic diffeomorphism invariance of Einstein gravity seems to require string theory for its UV completion, the holographic dual to a generic NR QFT seems to simply be a UV fixed point of Ho\v rava gravity, with non-trivial dynamical scaling exponent $z_{HL}$, coupled to an almost arbitrary matter sector. This basic picture has been one of the motivations behind the original work of \cite{Horava:2008ih,Horava:2009uw} and was also recently emphasized in \cite{Griffin:2011xs,Berglund:2012fk}.

The organization of this paper is as follows: Section \ref{sec:NRQFT} discusses non-relativistic quantum field theories; focussing on their symmetry properties in \ref{sec:NRdiff}, including a form of conformal invariance in \ref{sec:FTconf}, as well as deriving their transformations, from a $c\to\infty$ limit of a relativistic field theory with a chemical potential set to compensate the rest mass, in \ref{sec:relpar}. Section \ref{sec:Hgravity} discusses the non-relativistic gravity theory of Ho\v rava; a version with the same symmetries as a generic NR QFT is developed in \ref{sec:veckhron}, and a holographic duality is proposed in section \ref{sec:hmap}, including the calculation of a correlation function in \ref{sec:corrfunc}. We end with a discussion of our results in section \ref{sec:discussion}, where we also elaborate on string theory embeddings of our construction. A brief summary of the main ideas in this paper appears elsewhere \cite{short}.

\section{Field theories with non-relativistic electro-diffeomorphism invariance}\label{sec:NRQFT}
\subsection{Diffeomorphisms and the global $U(1)$ symmetry}\label{sec:NRdiff}

As first introduced in \cite{Son:2005rv}, and extended in \cite{Son:2008ye}, many NR QFTs with conserved particle number are invariant under diffeomorphism and $U(1)$ transformations if the background fields transform as
\begin{eqnarray}
 \nonumber
\delta A_t &=& \xi^\mu\d_\mu A_t+\dot{f}A_t+A_k\dot{\xi}^k-\dot{\lambda},\\
\nonumber
\delta A_i&=& \xi^\mu\d_\mu A_i+A_k\d_i\xi^k+A_t\d_i f+m e^\Phi g_{ik}\dot{\xi}^k-\d_i\lambda,\\
\nonumber
\delta\Phi&=& \xi^\mu\d_\mu\Phi+B_k\dot{\xi}^k-\dot{f},\\
\nonumber
\delta B_i&=&\xi^\mu \d_\mu B_i+B_k\d_i \xi^k+B_i(B_k\dot{\xi}^k-\dot{f})-\d_i f,\\
\label{QFTtrafo}
\delta g_{ij}&=& \xi^\mu \d_\mu g_{ij}+g_{ik}\d_j\xi^k+g_{kj}\d_i\xi^k+(B_ig_{jk}+B_jg_{ik})\dot{\xi}^k.
\end{eqnarray}
The diffeomorphism parameters, $\xi^t\equiv f$ and $\xi^i$, and the gauge parameter $\lambda$ can be arbitrary functions of space and time. This is the symmetry group of ``NR electro-diffeomorphisms" defined in the introduction. We can give an interpretation to these background fields by examining an action with this symmetry. Consider free NR particles described by the action
\begin{eqnarray}
 \nonumber
 S=\int dt d^dx\sqrt{g}e^{-\Phi}&&\left[\frac{\imath}{2}e^\Phi(\psi^\dagger D_t\psi-D_t\psi^\dagger\psi)-\frac{g^{ij}}{2m} D_i\psi^\dagger D_j\psi \right. \\
\label{NRaction}
&&\left. -\frac{g^{ij} B_j}{2m} (D_t\psi^\dagger D_i\psi+D_i\psi^\dagger D_t\psi)-\frac{g^{ij}B_iB_j}{2m}D_t\psi^\dagger D_t\psi\right],
\end{eqnarray}
where $D_\mu\psi\equiv\d_\mu\psi-\imath A_\mu \psi$ is the gauge covariant derivative. This action is invariant under the transformations \ref{QFTtrafo} if the field $\psi$ transforms as
\begin{equation}
 \delta\psi=\xi^\mu\d_\mu\psi-\imath \lambda \psi.
\end{equation}
By varying the action \ref{NRaction} with respect to the background fields we can give them physical meaning \cite{Son:2008ye}: $g_{ij}$ is the spatial metric and couples to the stress tensor $T^{ij}$; $A_\mu$ is the gauge field and couples to the particle number density and current $(n,\vec{j})$; and $(\Phi,\vec{B})$ are the sources that couple to the energy density and current $(\epsilon,\vec{E})$.

Among the general transformations described by \ref{QFTtrafo} is a subgroup that leaves the trivial background, $g_{ij}=\delta_{ij}$ and $A_\mu=\Phi=B_i=0$, invariant. These are determined to be translations, spatial rotations, and Galilean boosts. The latter are given by
\be
\vec{\xi}(t,\vec{x})=\vec{v}t,\qquad\lambda(t,\vec{x})=\vec{v}\cdot \vec{x}.
\ee
In this sense the only true non-trivial symmetry that is a consequence of NR electro-diffeo\-morphism invariance are Galilean boosts. More general transformations are only a symmetry if we treat the background fields as spurions, transforming according to \ref{QFTtrafo}.

\subsection{Conservation laws}
The spurionic symmetry transformations of the background fields, as captured in \ref{QFTtrafo}, leads to expressions for the conservation of particle number, momentum, and energy \cite{Son:2005rv,Son:2008ye}. In general backgrounds the latter two are only conserved if one takes into account the momenta and energy stored in the external fields. The connected part of the generating functional, $W$, is defined as $e^{\imath W}\equiv \int D\psi^\dagger D\psi e^{\imath S}$. Assuming that $W$ can be written as an integral of a local density,
\begin{equation}
 W[\Phi,B_i,g_{ij},A_t,A_i]=\int dt d^dx \mathcal{W},
\end{equation}
the invariance of the action $S$ under the field transformations \ref{QFTtrafo} implies, upon integrating by parts, the conservation laws:
\begin{equation}
 \d_tn+\d_kj^k=0,\quad \d_t\pi_i+\d_kT^k_i=0,\quad \d_t\epsilon+\d_kE^k=0,
\end{equation}
which are the conservation of particle number, momentum, and energy, respectively. The conserved densities and currents are given by
\begin{eqnarray}
 \nn n &\equiv& -\frac{\delta \mathcal{W}}{\delta A_t},\qquad j^k\equiv -\frac{\delta \mathcal{W}}{\delta A_k},\\
\nn \pi_i &\equiv& -B_i\left(\frac{\delta \mathcal{W}}{\delta \Phi}+B_j\frac{\delta \mathcal{W}}{\delta B_j}\right) - \left(B_kg_{ij}+B_jg_{ik}\right)\frac{\delta \mathcal{W}}{\delta g_{kj}}-A_i \frac{\delta \mathcal{W}}{\delta A_t}-m e^\Phi g_{ij}\frac{\delta \mathcal{W}}{\delta A_j},\\
\nn T^k_i &\equiv& \delta^k_i\mathcal{W}-B_i\frac{\delta \mathcal{W}}{\delta B_k}-2g_{ij}\frac{\delta \mathcal{W}}{\delta g{kj}}+A_i\frac{\delta \mathcal{W}}{\delta A_k}, \\
\epsilon &\equiv& \frac{\delta \mathcal{W}}{\delta \Phi}+B_i\frac{\delta \mathcal{W}}{\delta B_i}-A_t\frac{\delta \mathcal{W}}{\delta A_t},\qquad E^k\equiv\frac{\delta \mathcal{W}}{\delta B_k}-A_t\frac{\delta \mathcal{W}}{\delta A_k}.
\end{eqnarray}

In what follows we will be mostly interested in the case of $B_i=0$. For such backgrounds the momentum density and particle number current are linked \cite{Son:2005rv,Greiter:1989qb}
\be
\pi_i=nA_i+m e^\Phi j_i.
\ee
Even when $\Phi=B_i=0$ the variation of $\mathcal{W}$ with respect to these fields is needed to to calculate the energy density $\epsilon$ and current $\vec{E}$.

\subsection{NR scale and conformal invariance}\label{sec:FTconf}
In addition to the above diffeomorphism and $U(1)$ transformations we can extend the spurionic symmetry of some NR QFTs to include a type of conformal invariance \cite{Son:2008ye}. The additional generator $\omega(t,\vec{x})$ acts on the background fields via
\begin{eqnarray}
\label{conf}
\delta_\omega \Phi &=&-2\omega, \qquad \delta_\omega g_{ij} = 2\omega g_{ij},
\end{eqnarray}
with the rest invariant.

Although the action \ref{NRaction} is not invariant under this transformation, it can be made so by exchanging the ``minimal coupling'' used here for ``conformal coupling'' \cite{Son:2008ye}. Alternatively, we can consider the restricted case of $\Phi=B_i=0$, as in \cite{Son:2005rv}. To maintain $B_i=0$, from \ref{QFTtrafo}, we require $\d_i f=0$, which is simply the statement that non-physical temporal diffeomorphisms are not allowed in NR general covariance. Conversely, the existence of a global time allows\footnote{On the spatial leaves defined by a global time we require that the above action \ref{NRaction} reproduces the Schr\"odinger equation. This in turn gives $B_i=0$ in such a coordinate frame. Using the observation that $n_\mu\equiv(e^{-\Phi},-e^{-\Phi}B_i)$ transforms as a spacetime one-form, we can interpret the action \ref{NRaction} as giving time evolution in the $n_\mu$ direction.} $B_i=0$, as long as we work in adapted
coordinates. From \ref{QFTtrafo}
 and \ref{conf} we see that $\Phi=0$ is maintained for $\omega=-\dot{f}(t)/2$. In this way we see how in the restricted case of \cite{Son:2005rv} time reparametrization contains the information of the conformal structure of the theory. They are intimately linked by demanding that $\Phi$ remains zero.

It is useful to define the notion of the conformal dimension of an operator \cite{Son:2005rv}. By the argument above, we see that for NR general covariance this can be determined by the operator's behavior under infinitesimal time reparametrization. In general an operator/field transforming as
\be
\delta \mathcal{O}\supset f\dot{\mathcal{O}}+\Delta_\mathcal{O}\dot{f}\mathcal{O}
\ee
is said to have the conformal dimension $\Delta_\mathcal{O}$. From \ref{QFTtrafo} and \ref{conf}, and using $\omega=-\dot{f}/2$, we see that in the restricted case the remaining background fields transform as
\begin{eqnarray}\label{confdim}
\delta A_t\supset f\dot{A}_0+\dot{f}A_t,\quad\delta A_i\supset f \dot{A}_i,\quad \delta g_{ij}\supset f\dot{g}_{ij}+2\omega g_{ij}=f\dot{g}_{ij}-\dot{f}g_{ij}.
\end{eqnarray}
Therefore, $A_t$, $A_i$, and $g_{ij}$ are conformal operators with dimensions $1$, $0$, and $-1$, respectively. With these transformations of the background fields the action \ref{NRaction} (with $\Phi=B_i=0$) is invariant under arbitrary $f(t)$; the free action is ``conformally invariant'' if we assign the scalar field the conformal transformation
\be
\delta \psi\supset -\frac{d}{2}\omega\psi=\frac{d}{4}\dot{f}\psi.
\ee

If we formulate a theory with this conformal invariance there is a subgroup of the spurionic symmetry transformations that leave the trivial background $g_{ij}=\delta_{ij}$ and $A_\mu=0$ invariant. We have already seen that translations, rotations, and Galilean boosts maintain this background. A second special case is the scale transformation. This corresponds to a constant conformal transformation, $\omega=-\kappa/2$, which, by above, requires the time reparametrization $f=\kappa t$. In order to leave the trivial background metric invariant we need to combine these transformations with a spatial diffeomorphism that corresponds to rescaling the spatial coordinates
\be
\xi^i=\frac{\dot{f}}{2}x^i=\frac{\kappa}{2}x^i.
\ee
The relative weight of $1/2$ between the rescaling of time and space corresponds to a dynamical critical exponent of $z=2$, as expected for a Schr\"odinger system.

In later sections we will find examples of gravity backgrounds that have scaling symmetries for $z\ne2$. In order for this more general scale transformation to be a symmetry of a Galilean invariant QFT the conformal transformation must be modified to
\be\label{zne2}
\delta_\omega \Phi =-z\omega, \qquad \delta_\omega g_{ij} = 2\omega g_{ij},\qquad \delta_\omega m = (z-2)\omega m,
\ee
that is, the mass $m$ must now be treated as a spurionic field. Preserving the trivial background under the temporal rescaling $f=\kappa t$ then requires the conformal transformation $\omega=-\kappa/z$ and the spatial rescaling $\xi^i=\kappa x^i/z$, as expected for dynamical critical exponent $z$. It has been argued in \cite{Balasubramanian:2008dm}, based on a holographic construction, that such scale and Galilean invariant fixed points should exist in interacting NR QFTs. In the action \ref{NRaction}, as in Schr\"odinger's equation, $m$ is a parameter, not a dynamical field. In this case, $z=2$ scale transformations get singled out as the only true scale symmetry that leaves the mass invariant. All other values of $z$ can formally be realized as spurionic symmetries under which $m$ transforms. This is also the case in the $z \neq 2$ backgrounds of \cite{Balasubramanian:2008dm}, where the compact light-like direction scales non-trivially for $z\neq 2$, and hence so does the compactification radius which sets
the mass of the Kaluza-Klein particles. In principle we can construct a system with $z \neq 2$ scaling by promoting $m$ to a dynamical field in all the above, and adding a hidden sector action $S_m$ which sets the scaling of $m$ to be given by \ref{zne2}. We will not attempt to construct an explicit field theory model that realizes such behavior.

For the $z=2$ case realized by the free field theory above, there is one more symmetry generator that leaves the trivial background invariant. It is usually referred to as the ``special conformal'' transformation of the Schr\"odinger group, and corresponds to the combination
\be
\omega= -Ct, \quad f=Ct^2,\quad\xi^i=Ctx^i,\quad\lambda=\frac{1}{2}C\vec{x}^2.
\ee

Interactions can be added to the free theory that preserve spatial diffeomorphism and the global $U(1)$ invariance. In particular, the physically important case of a Coulomb interaction (as relevant for electron systems) has the full NR general covariance, while a short range interaction in the limit of infinite scattering length (as relevant for the unitary Fermi gas) additionally has NR conformal invariance. Hence their low energy physics are constrained by these symmetries. General interactions need not preserve the full conformal symmetry of the free theory. If the theory remains invariant under transformations of the form $f(t)=f_0+f_1t$ then it has time translation and scale invariance. For the case of $f_1=0$ the theory only has time translation invariance.

\subsection{Relativistic parent theory}\label{sec:relpar}
The transformations \ref{QFTtrafo} can easily be derived by taking a non-relativistic limit of the relativistic theory of a charged massive field. As mentioned in the introduction, this can be achieved by introducing a chemical potential to supply the rest mass of particles, and then taking the $c\to\infty$ limit to focus on fluctuations around this energy. Of course this procedure does not give the most general NR QFT, but it does give a simple way to derive the transformation properties of the free field theory. This is easiest to illustrate in the case of a scalar \cite{Son:2005rv}. The relativistic action
\be\label{Raction}
S=-\int d^dxdt\sqrt{-g}\frac{1}{2}\left(g^{\mu\nu}\mathcal{D}_\mu\phi^\dagger\mathcal{D}_\nu\phi+c^2m^2e^{2\sigma}\phi^\dagger\phi\right),
\ee
where we have introduced the gauge covariant derivative $\mathcal{D}_\mu\phi\equiv\d_\mu\phi-\imath C_\mu\phi$, is invariant under the infinitesimal general relativistic coordinate and $U(1)$ gauge transformations
\begin{eqnarray}
 \nonumber
\delta \phi&=&\xi^\rho\d_\rho\phi-\imath\Lambda\phi,\\
\nonumber
\delta C_\mu&=&\xi^\rho\d_\rho C_\mu+C_\rho\d_\mu\xi^\rho-\d_\mu\Lambda,\\
\label{Rtrafo}
\delta g_{\mu\nu}&=&\xi^\rho\d_\rho g_{\mu\nu}+g_{\mu\rho}\d_\nu\xi^\rho+g_{\rho\nu}\d_\mu\xi^\rho.
\end{eqnarray}

We explicitly display powers of the speed of light in the action so we can take the non-relativistic $c\to\infty$ limit. Note that the relativistic mass is defined as $me^{\sigma}$, this is crucial as it has a different scaling dimension than the non-relativistic mass $m$, as discussed above. Additionally, we will allow $m$ and $\sigma$ to have spacetime dependence. Following \cite{Son:2005rv} we would now define the non-relativistic field by factoring out the fast phase rotation due to the scalar field's rest mass: $\phi\equiv e^{-\imath m c^2 t}\phi_{NR}/\sqrt{c}$. For our charged scalar we can instead gauge away this phase via the gauge transformation $\Lambda = -c^2mt$. Therefore we can treat $\sqrt{c}\phi$ itself as a non-relativistic charged scalar by working with the background $C_\mu=-\d_\mu \Lambda = \delta_{\mu t}mc^2$.  Although relativistically such a gauge field would be considered highly trivial as it has zero field strength, here it plays an important role due to the fact that $c$ dependent
gauge transformations, such as the above $\Lambda$, and the $c\to\infty$ limit do not commute. Also note that, unlike constant spatial vector potentials a constant $C_t$ can, in general, not be completely gauged away.
The term $\int_M j^{\mu} C_{\mu}$ (where $M$ is the space-time manifold) is usually taken to be gauge invariant as long as $j^\mu$ is a conserved current. Under a gauge transformation $\delta C = -d \Lambda$, the change in action is
\be
\delta S = -\int_M j^{\mu} \partial_{\mu} \Lambda = - \int_{\partial M} (\Lambda j_{\mu}) dS^{\mu} + \int_M \Lambda \partial_{\mu} j^{\mu}.
\ee
The second term vanishes by current conservation. The contributions to the boundary term from spatial boundaries vanish for any localized current. However, for the boundaries of the integral at the final and initial times, $t=t_f$ and $t=t_i$, we can not take $j^0$ to vanish. The total charge $Q$ is conserved, and if it is non-zero at one time, it is non-zero at all times. In particular, for $\Lambda=m c^2 t$ (which would be needed to set our constant $C_t$ to zero) one has
\be
\delta S =\left .  -mc^2 Q t \right |^{t_f}_{t_i} = mc^2 Q (t_i-t_f)
\ee
which clearly is non-zero as long as $Q$ is non-zero. The action is only invariant under the restricted class of gauge transformations which vanish at $t_f$ and $t_i$. However, to remove a constant $C_t$ would require a gauge transformation which is non-vanishing on initial and final surfaces.

As a warm-up, consider, as in \cite{Son:2005rv}, the metric expansion\footnote{The leading piece of $g_{tt}$ goes as $c^2$ as we are using the non-relativistic time $t$ as our temporal coordinate, not $x^0\equiv ct$. Likewise for the behavior of $g_{ti}$.}
\begin{equation}
g_{\mu\nu}=
 \begin{pmatrix}
  -c^2+2\frac{A_t}{m} & \quad \frac{A_i}{m}\\
\frac{A_j}{m} & \quad g_{ij}
 \end{pmatrix}.
\end{equation}
For a constant $m$ and $\sigma=0$, plugging this form of the metric, the gauge field background $C_t=mc^2$, and the rescaled field $\psi=\sqrt{mc}\phi$ into the relativistic action \ref{Raction}, and after discarding negative powers of $c$, we obtain
\be\label{swaction}
S=\int d^dxdt\sqrt{g}\left[\frac{\imath}{2}(\psi^\dagger\d_t\psi-\d_t\psi^\dagger\psi)+A_t\psi^\dagger\psi-\frac{g^{ij}}{2m}(D_i\psi^\dagger D_j\psi)\right],
\ee
which is the non-relativistic action \ref{NRaction} with $\Phi=B_i=0$. The action of spatial diffeomorphisms and the global $U(1)$ on the remaining background fields can be determined from \ref{Rtrafo} for the generators $\xi^\mu=(\lambda/mc^2,\xi^i)$.

There are two important points in the details of this calculation, both concerning the gauge field $C_\mu$. First, the mass term in the relativistic action \ref{Raction}, for $\sigma=0$, would contribute the term $-1/2 c^2m^2\psi^\dagger\psi$ to the non-relativistic action, forcing $\psi=0$ in the $c\to\infty$ limit. For the background $C_t=mc^2$ this mass term is canceled by the $-C_\mu g^{\mu\nu}C_\nu\psi^\dagger\psi$ term coming from the covariant derivative. This is understood as tuning a chemical potential that provides the rest mass of the particles, so that the non-relativistic action only describes fluctuations around this energy. Thus the magnitude of the gauge field acts as a chemical potential and needs to be fixed to cancel the mass term and allow a non-trivial non-relativistic limit.

Secondly, we need to be assured that we have done a consistent expansion in powers of $c$, both of the metric and the gauge field. We see that $C_\mu$ naturally has a piece of order $c^2$ from performing the gauge transformation removing the fast phase of the scalar field. A consistent expansion can be made so that the next term comes in at zeroth order: $C_\mu=c^2b_\mu+v_\mu+\mathcal{O}(c^{-2})$. As will be discussed in more detail in sections \ref{sec:EMFT} and \ref{sec:alpha}, for the expansion of the temporal diffeomorphism generator $\xi^t=f-\alpha/c^2$ the gauge field transforms as
\be
\delta v_\mu\supset -C_t\d_\mu\alpha/c^2=-b_t\d_\mu\alpha,
\ee
that is, the $\mathcal{O}(c^0)$ piece of $C_\mu$ is generated by the subleading temporal diffeomorphism $\alpha$. As the warm-up example above had $v_\mu=0$ throughout, to maintain this restriction we implicitly performed an $\mathcal{O}(c^0)$ gauge transformation. Therefore, the appearance of the gauge transformation $\lambda$ as the subleading term of the temporal diffeomorphism is only an artifact of demanding $\alpha=-\lambda/m$, so that $v_\mu$ stayed zero.

We would like to extend the warm-up example to include general backgrounds. As discussed above, the gauge field can be consistently expanded as $C_\mu\equiv c^2b_\mu+v_\mu$. We can determine the consistent expansion of the metric by first considering the case of $C_\mu=c^2\delta_{\mu}^t b_t$, similar to the warm-up example. In this frame we parametrize the metric in the ADM form as\footnote{The lack of $A_t$ in this expansion will be discussed further in section \ref{sec:bulkact}.}
\begin{equation}
g_{\mu\nu}=
 \begin{pmatrix}
  -c^2 N^2+N^kN_k & \quad N_i\\
N_j & \quad G_{ij}
 \end{pmatrix},
\end{equation}
where $N^k=G^{ki}N_i$. The general leading gauge field $C_\mu=c^2b_\mu$ can be obtained from $C_t=c^2b_t$ by performing a coordinate transformation, $C_\mu\to J^\nu_\mu C_\nu$. Under this transformation the metric change by two Jacobian factors, $g_{\mu\nu}\to J^\rho_\mu g_{\rho\sigma}J^\sigma_\nu$, and it can be see that all components generically have $\mathcal{O}(c^2)$ pieces. We are therefore lead to expand the metric as\footnote{This can be seen to be a consistent expansion, meaning no other positive power of $c$ pieces get turned on by coordinate transformations.}
\begin{equation}\label{genmetric}
g_{\mu\nu}=
 \begin{pmatrix}
  -c^2 N^2+N^kN_k & \quad N_i+c^2P_i\\
N_j+c^2P_j & \quad G_{ij}-c^2\frac{P_iP_j}{N^2}
 \end{pmatrix}.
\end{equation}

We are now in position to derive the non-relativistic action \ref{NRaction} and transformations \ref{QFTtrafo} by taking the formal $c\to\infty$ limit of the relativistic theory. First we must be sure that the chemical potential provides the rest energy of our particles. As discussed above this is achieved by a cancellation between the mass term and the magnitude of the gauge field. In general there are $\mathcal{O}(c^4)$ pieces of $C_\mu C^\mu$. For the action to have a non-trivial non-relativistic limit this piece must vanish, requiring
\be \label{pibi} \frac{P_i}{N^2}=-\frac{b_i}{b_t}.\ee
This can be understood as the requirement that the theory has a global time, needed for a causal non-relativistic theory. As discussed above in section \ref{sec:FTconf}, in adapted coordinates the vector $B_i$ vanishes. The relation of this NR field to $b_i$ and $P_i$, given in \ref{RNRdict} below, justifies the identification \ref{pibi}: in general these fields only arise due to coordinate changes to a non-adapted frame. The $\mathcal{O}(c^2)$ piece of $C_\mu C^\mu$ will play the role of a chemical potential and cancel the mass term. Explicitly this requires
\be\label{sigmadep}
\frac{b_t}{N}= m e^{\sigma}.
\ee
Plugging the rescaled field $\phi\to\phi/\sqrt{c}$ and our expansions into the relativistic action \ref{Raction} (with the generalized spacetime dependent mass), and discarding negative powers of $c$, we obtain
\be
S=\int dt d^dxme^\sigma \mathcal{L},
\ee
where $\mathcal{L}$ is the Lagrangian density of \ref{NRaction} if we make the identifications
\begin{eqnarray}
\nonumber
 e^{-\Phi}&\equiv&\frac{m N^2}{b_t},\\
\nonumber
B_i&\equiv&-\frac{b_i}{b_t}=\frac{P_i}{N^2},\\
\nonumber
A_t &\equiv& v_t+\frac{b_tN^kN_k}{2N^2},\\
\nonumber
A_i&\equiv& v_i+\frac{b_tN_i}{N^2}-\frac{b_iN^kN_k}{2N^2},\\
\label{RNRdict}
g_{ij}&\equiv& G_{ij}-\frac{b_iN_j}{b_t}-\frac{b_jN_i}{b_t}+\frac{b_ib_jN^kN_k}{b_t^2}.
\end{eqnarray}

These field combinations transform as \ref{QFTtrafo} if we expand the relativistic generators of \ref{Rtrafo} as $\xi^\mu=(f,\xi^i)$ and $\Lambda=\lambda$. We can no longer do $\mathcal{O}(c^2)$ gauge transformations as they would not leave the $\mathcal{O}(c^2)$ piece of $C_\mu C^\mu$ invariant, which was needed to have a well defined $c\to\infty$ limit. The role of the subleading temporal diffeomorphism $\alpha$, introduced previously, and its relation to the specific combinations of relativistic fields in \ref{RNRdict} will be discussed in section \ref{sec:alpha}.

For constant $m$ and $\sigma$ the field can be rescaled $\psi\equiv\phi\sqrt{m}e^{\sigma/2}$ and the above exactly reproduces the Lagrangian density of \ref{NRaction}. This rescaling only changes the dimension of the field and, in fact, can be done even for time dependent $m$ and $\sigma$. We can most easily understand the role of the field $\sigma$ by enforcing NR conformal invariance on the above action. For the general $z\ne 2$, the transformation \ref{zne2} will be a spurionic symmetry of the action if
\begin{equation}
\label{spurionicomega}
 \delta_\omega e^\sigma = -(d+z-2)\omega e^\sigma.
\end{equation}
It is now clear why the NR action \ref{swaction} of \cite{Son:2005rv} has NR conformal invariance. One can use the transformation $\omega$ to set $me^\sigma$ to a constant. We then consider the restricted case of \cite{Son:2005rv} with $B_i=\Phi=0$. This is maintained by performing $\omega=-\dot{f}/z$ whenever the temporal redefinition $f(t)$ is performed. In turn, $me^{\sigma}$ will generically become a function of time. But such a factor can be absorbed into the fields by a redefinition, due to the anti-symmetric nature of the time derivative in \ref{NRaction}. This means the restricted case \cite{Son:2005rv} will have NR conformal invariance and the field has the transformation
\be
\delta_\omega \psi \equiv \delta_\omega \left(\phi\sqrt{m}e^{\sigma/2}\right)=\frac{1}{2\sqrt{m}}\phi e^{\sigma/2}\delta_\omega m+\frac{1}{2} \phi \sqrt{m}e^{-\sigma/2} \delta_\omega e^\sigma=-\frac{d}{2}\omega \psi.
\ee

\section{Gravitational theories with non-relativistic general covariance}\label{sec:Hgravity}
\subsection{Ho\v rava gravity}
\subsubsection{Foliation preserving diffeomorphisms}
Ho\v rava gravity \cite{Horava:2009uw} is a metric theory built around foliation preserving diffeomorphisms, that is, time dependent spatial diffeomorphisms and time reparametrization. The minimal set of fields common to all versions of Ho\v rava gravity are the lapse $N(t,x_I)$, the shift vector $N_I(t,x_I)$, and the spatial metric $G_{IJ}(t,x_I)$. In the projectable version of the theory one would restrict $N$ to be a function of $t$ only, but we will not do so here. The action can be written as
\begin{equation}\label{Haction}
 S=\int dt d^dx dr\left(\mathcal{L}_{kin}-\mathcal{L}_V\right),
\end{equation}
where the kinetic term is given in terms of the extrinsic curvature of the leaves,
\begin{equation}
 K_{IJ}\equiv \frac{1}{2N}\left(\dot{G}_{IJ}-\nabla_I N_J-\nabla_J N_I\right),
\end{equation}
and its trace, $K=G^{IJ}K_{IJ}$, by
\begin{equation}
 \mathcal{L}_{kin}=\frac{1}{16\pi G_N}\sqrt{G}N\left[K_{IJ}K^{IJ}-\left(\tilde{\lambda}+1\right) K^2\right].
\end{equation}
Here $G$ is the determinant of the spatial metric $G_{IJ}$, and $\nabla_I$ is its Levi-Civita connection. The simplest potential term involving up to two derivatives, as appropriate for the low energy or large $N_c$ limit, is given by \cite{Blas:2009yd,Blas:2010hb}
\begin{equation}
 -\mathcal{L}_V=\frac{1}{16\pi G_N}\sqrt{G}N\left[R-2\Lambda+\alpha\frac{(\nabla_I N)(\nabla^I N)}{N^2}\right],
\end{equation}
where $R$ is the Ricci scalar of $G_{IJ}$ and $\Lambda$ is the cosmological constant.

The constants $\tilde{\lambda}$ and $\alpha$ are free dimensionless coupling constants that are allowed by demanding only foliation preserving diffeomorphisms and not the full relativistic diffeomorphism invariance of GR. For $\tilde{\lambda} =\alpha=0$ this becomes, up to a total derivative, the standard Einstein-Hilbert action written in terms of the ADM decomposition of the full $d+2$ dimensional bulk metric
\begin{equation}
 \tilde{G}_{MN}=
\begin{pmatrix}
-N^2+N^KN_K & \quad N_I\\
N_J & \quad G_{IJ}
 \end{pmatrix}.
\end{equation}
Even in the $\tilde{\lambda}=\alpha=0$ limit this is not the theory of standard GR. Despite identical actions, the gauge invariances of Ho\v rava gravity lack the general temporal diffeomorphism $t\to \tilde{t}(t,x_I)$. As a consequence Ho\v rava gravity contains an extra scalar degree of freedom as compared to GR.

Under spatial diffeomorphisms $\xi^I$ and time reparametrizations $f(t)$ the fields transform as
\begin{eqnarray}\nn
 \delta G_{IJ}&=&\xi^K\d_K G_{IJ}+f\dot{G}_{IJ}+G_{IK}\d_J\xi^K+G_{KJ}\d_I\xi^K,\\
\nn\delta N_I&=&\xi^K\d_K N_I+f\dot{N}_I+N_K\d_I\xi^K+G_{IK}\dot{\xi}^K+\dot{f}N_I,\\
\label{Htrans}\delta N&=&\xi^K\d_K N+f\dot{N}+\dot{f}N.
\end{eqnarray}
These can be derived by taking the $c\to\infty$ limit of the transformation of the relativistic metric $\tilde{G}_{MN}$ with the diffeomorphism parameters $\xi^M=\left(f,\xi^I\right)$ (after explicitly restoring the speed of light to the metric: $N\to cN$) \cite{Horava:2008ih}.

\subsubsection{NR general covariance and the scalar khronon}\label{sec:scalar}
Ho\v rava gravity can be usefully embedded into standard GR via a St\"uckelberg-like mechanism \cite{Germani:2009yt,Blas:2009yd,Blas:2010hb}. This formalism makes the extra degree of freedom explicit by coupling Einstein gravity to an additional scalar field $\phi$. When the scalar field acquires an expectation value $\phi=t$ the symmetry of GR is broken down to only spatial diffeomorphisms along the level sets of $\phi$. In this way $\phi$ can be used to define the preferred foliation by a global time, and is referred to as the khronon \cite{Blas:2009yd,Blas:2010hb}. This view of Ho\v rava gravity as GR with diffeomorphism invariance broken by a background field has also been recently emphasized in \cite{Kiritsis:2012ta}.

To have the symmetries of foliation preserving diffeomorphisms, $\phi$ needs to have the reparametrization symmetry in field space $\phi\to\tilde{\phi}(\phi)$, which becomes the time reparametrization symmetry of
Ho\v rava gravity\footnote{As
explained in \cite{Blas:2010hb}, a similar construction also underlies other modified theories of gravity. A time dependent condensate of a scalar with a shift symmetry (giving rise to a theory with time dependent spatial diffeomorphisms together with time translation symmetry) underlies the ``ghost condensation" model \cite{ArkaniHamed:2003uy} as well as shift-symmetric $k$-essence \cite{ArmendarizPicon:1999rj}. When even time translation symmetry is absent and only time dependent spatial diffeomorphisms are preserved, the symmetry group governs the effective theory of standard inflation \cite{Cheung:2007st,Weinberg:2008hq}. If time translation invariance is combined with time independent diffeomorphisms one has the symmetry of Einstein-aether theory \cite{Jacobson:2000xp} or gauged ghost
condensation \cite{Cheng:2006us}.}. This reparametrization invariance can be made explicit by working with the time-like unit vector normal to the leaves of constant $\phi$,
\be u_M\equiv\frac{-\d_M\phi}{\sqrt{-\tilde{G}^{NP}\d_N\phi\d_P\phi}}. \ee
In the ``unitary gauge,'' where we choose our time coordinate to be the expectation value of the khronon, $t=\phi$, we have $u_0=-N$ and all the spatial components vanish. The geometric quantities of the foliation appearing in Ho\v rava gravity can all be expressed in terms of the khronon field. In particular, in unitary gauge the spatial components of
\be
\mathcal{K}_{MN}\equiv\left(\tilde{G}_{MP}+u_Mu_P\right)\tilde{\nabla}^Pu_N
\ee
become the extrinsic curvature $K_{IJ}$. The Ho\v rava action \ref{Haction} can now be written as the EH action coupled to the scalar khronon \cite{Blas:2010hb}
\be
S_{kh}=\frac{1}{16\pi G_N}\int dtd^dxdr\sqrt{-\tilde{G}}\left[\tilde{R}-2\Lambda+\tilde{\lambda}\left(\tilde{\nabla}_Mu^M\right)^2+\alpha\left(u^M\tilde{\nabla}_Mu^P\right)\left(u^N\tilde{\nabla}_Nu^P\right)\right].
\ee
The reader should recall that the tilded quantities refer to those derived from the full $d+2$ dimensional Lorentzian metric.

A powerful use of the khronon formalism is that for $\tilde{\lambda}$ and $\alpha$ parametrically small we can treat $\phi$ as a probe field since its stress tensor does not backreact on the metric. One can then solve the full non-linear gravitational equations of motion of Ho\v rava gravity by starting with a solution to Einstein gravity and solving for the khronon field on this background. A non-trivial khronon field configuration can then be reinterpreted in unitary gauge as a solution to Ho\v rava gravity \cite{Blas:2011ni}. Concretely, once we find the solution $\phi=t+\chi(t,x_I)$ in a given GR background we perform the relativistic diffeomorphism $\tilde{t}=t+\chi(t,x_I)$ to go to unitary gauge. The resulting lapse, shift, and spatial metric is now a solution to Ho\v rava gravity.

One can also use the scalar khronon to formulate the generally covariant version of Ho\v rava gravity \cite{Horava:2010zj,daSilva:2010bm}. As initially introduced in \cite{Horava:2008ih} the transformations of the Ho\v rava fields \ref{Htrans} can be extended to include a $U(1)$ transformation by expanding to the next order in the speed of light. For $N\to N + A(t,x_I)/c^2$ and $\xi^t=f-\alpha(t,x_I)/c^2$ the action of the $U(1)$ transformation $\alpha$ is
\begin{eqnarray}
 \nn \delta_\alpha N &=& 0,\quad\delta_\alpha G_{IJ}=0,\\
\nn \delta_\alpha N_I &=& N^2\d_I\alpha,\\
\label{alphatrans}\delta_\alpha A &=& -\dot{\left(\alpha N\right)}+NN^I\d_I\alpha,
\end{eqnarray}
while under foliation preserving diffeomorphisms $A$ transforms as $N$ does. As it stands the action \ref{Haction} is not invariant under this transformation. As first developed in \cite{Horava:2010zj}, and later generalized in \cite{daSilva:2010bm}, this can be fixed by postulating the ``pre-potential'' field $\nu$ that shifts under the $\alpha$ transformation. We will now show that this field can be associated with the scalar khronon.

For a consistent interpretation of $\alpha$ as a gauge transformation we need to understand how it acts on the khronon. Restoring factors of the speed of light we have
\be
\phi=c^2t+\chi(t,x_I)
\ee
for the expansion of the khronon around unitary gauge. From this we expect the transformation $t\to t+\alpha/c^2$ to be reinterpreted as the shift $\chi\to\chi-\alpha$, that is, the subleading relativistic temporal diffeomorphism $\alpha$ can be interpreted in a non-relativistic foliation preserving way as instead shifting the khronon fluctuation $\chi$. Therefore, we see that the pre-potential $\nu$ is naturally identified with $\chi$, the subleading piece of the khronon in the $c\to \infty$ expansion. The transformation of $\chi$ can also be found by considering the khronon to be the phase of a complex scalar. Expanding the relativistic transformation of a scalar, and demanding the reparametrization invariance of the khronon field, one finds
\be
\delta \chi = \xi^K\d_K\chi+f\dot{\chi}-\dot{f}\chi-\alpha.
\ee

It is easy to check that the following combinations are invariant under the $U(1)$ transformation $\alpha$
\be \hat{N}_I\equiv N_I+N^2\d_I\chi,\quad\hat{A}\equiv A-\dot{(\chi N)}+NN^I\d_I\chi+\frac{N^3}{2}G^{IJ}\d_I\chi\d_J\chi. \ee
In the projectable case, this reproduces the ``minimal substitution'' of \cite{daSilva:2010bm} if we make the identification of the pre-potential with the khronon fluctuation\footnote{The factor of the lapse $N$ is due to our differing definition of $\alpha$ as the subleading piece of the temporal diffeomorphism when compared with \cite{Horava:2010zj,daSilva:2010bm}.}: $\nu\equiv -N\chi$. In particular our $\hat{A}$ is equivalent to \cite{daSilva:2010bm}'s $A-a$.

We can now understand an obstruction to using this form of covariant Ho\v rava gravity in a holographic duality. The khronon must be added to the bulk action to yield invariance under $\alpha$. From the statement of holography, this action can give the correlation function of the operator dual to the khronon by examining its on-shell boundary value. This operator is not gauge invariant though, and will shift under $\alpha$ as $\chi$ does. The only operator that shifts under a gauge transformation, that we are aware of, is the phase of a charged field; it acquiring a nontrivial correlation function indicates that the $U(1)$ symmetry is in fact broken in the field theory. This is also apparent by considering how the bulk transformation generated by $\alpha$ manifests itself as the global $U(1)$ rotation of the field theory. If one gauge fixes $N_r=0$, from \ref{alphatrans} it is seen that $r$ independent $\alpha$ maintains this bulk gauge choice and would be expected to correspond to boundary $U(1)$
transformations, leading to the above issue. One could
instead
use $\alpha$ to gauge away the khronon in the bulk. This does not solve the issue as now there are no residual $\alpha$ transformations that could be interpreted as acting on the boundary data. In this case we see the boundary $U(1)$ appears broken too.

There are two additional issues with the scalar khronon formulation leading us to abandon it as a holographic gravitational theory. First is a purely classical gravitational consideration. By its nature the khronon field needs a uniform spatial distribution to define the leaves of the foliation. Such a configuration should generically be gravitationally unstable to clumping, and therefore may not even define a consistent theory\footnote{Problems along these lines are known in the related ghost condensation theories \cite{Dubovsky:2004sg,Rubakov:2004eb}.}. The second issue is quantum in nature. In order to recover the time reparametrization invariance of Ho\v rava gravity the khronon $\phi$ needed to have a global field redefinition symmetry. In quantum gravity there is expected to be no global symmetries so this construction seems problematic beyond the classical level.

These shortcomings hint at a solution; as the khronon is seen to transform as the phase of a complex scalar, we should consider this scalar charged, and include the accompanying gauge fields in the bulk.  Being a gauged phase, this field would have no stress tensor and therefore avoid the issue of clumping. Time reparametrization can be implemented without the need of postulating global symmetries, and therefore can be consistent with tenets of quantum gravity. As this construction requires the inclusion of a bulk vector field to set a preferred time slicing we will refer to it as a vector khronon.
The hope of \cite{Horava:2010zj,daSilva:2010bm}, that the shift $N_I$ could play a dual role as a gauge field for both spatial diffeomorphisms $\xi^I$ and the $U(1)$ generator $\alpha$ seems to not be borne out, at least for holographic purposes. We will pursue the role of bulk gauge fields shortly, but first discuss an alternate motivation for their necessity.

\subsection{Vector khronons}\label{sec:veckhron}
\subsubsection{Kaluza-Klein vector khronon} \label{sec:kkk}
The first attempts \cite{Son:2008ye,Balasubramanian:2008dm,Goldberger:2008vg} at a gravitational dual to a non-relativistic field theory shared an unexpected feature: they had two extra dimensions compared to the NR QFT they described. This can be understood by realizing they are basically light-like compactifications of relativistic field theories in one higher dimension. With compactification on a light-like circle, the lower dimensional field theory preserves a non-relativistic subgroup of the higher dimensional relativistic Lorentz symmetry, the Schr\"odinger group. The holographic dual description correspondingly is also a light-like compactification of general relativity on AdS spacetime. Momentum modes along the light-like direction, $\zeta$, appear as separate conserved particle number sectors in the NR QFT, not as spatial momentum modes. This direction and the traditional holographic radial coordinate gives two extra dimensions to the bulk geometry. For an interesting non-relativistic interpretation of this geometry see \cite{Duval:2008jg}.

Near the boundary, $r=0$, the metric can be parametrized as \cite{Son:2008ye}\footnote{The $r^{-4}$ ``Lifshitz'' term in \cite{Son:2008ye,Balasubramanian:2008dm} is unimportant for our purposes. It is separately invariant under the symmetry transformations.
In the dual field theory, introduction of this extra term corresponds to imposing twisted boundary conditions for R-charged fields along the light-like circle \cite{Herzog:2008wg,Adams:2008wt,Maldacena:2008wh}. This twisting removes some of the zero modes on the circle and makes the field theory more tractable. To get a non-trivial field theory with the desired Schr\"odinger invariance it is not needed and the light-like circle compactification alone suffices.}
\begin{equation}\label{metricson}
 d\hat{s}^2=-\frac{2e^{-\Phi}}{m r^2}\left(dt-B_idx^i\right)\left(d\zeta-A_tdt-A_idx^i\right)+\frac{g_{ij}dx^idx^j+dr^2}{r^2}.
\end{equation}
The gauge $g_{\mu r}=g_{\zeta r}=0$ has been chosen, but this does not completely fix the diffeomorphisms of the theory. Under the residual transformations the fields parametrizing the metric transform exactly like the NR QFT fields \ref{QFTtrafo}, for $\xi^\zeta\equiv\lambda$.

The NR QFT described by GR on this background is highly constrained: most of its properties are inherited from the relativistic theory upon the light-like compactification. For $d=2$ it is known that the field theory is simply the discrete light cone quantization of $\mathcal{N}=4$ Super Yang-Mills theory in four spacetime dimensions \cite{Herzog:2008wg,Maldacena:2008wh}. Field theory properties, such as hydrodynamics and thermodynamics, follow from this relativistic reduction \cite{Herzog:2008wg,Rangamani:2008gi}. Here we use this known duality as a motivation: it has long been understood that a light-like compactification can be equivalent to a spatial compactification on a circle of vanishing radius, plus an appropriate boost \cite{Seiberg:1997ad,Sen:1997we}. We will perform a $c\to \infty$ scaling limit to make a spatial compactification light-like and recover the
metric \ref{metricson}.

This construction is equivalent to considering a chemical potential that provides the rest mass of the charged Kaluza-Klein momentum modes for a purely spatial circle and then taking the $c \rightarrow \infty$ limit, exactly as we did in our field theory construction of section \ref{sec:NRQFT}. This allows us to directly identify the correct bulk fields that map to the field theory sources of section \ref{sec:NRQFT}, as well as the bulk version of the constraint relating the chemical potential to the rest mass.

Consider a $d+3$ dimensional spacetime with metric $\hat{G}_{XY}$, and compactify along the last direction $\zeta$. The Kaluza-Klein decomposition of the metric is
\be
\hat{G}_{XY}\equiv L^2
\begin{pmatrix}
 \tilde{G}_{MN}+G_{\zeta\zeta}C_MC_N & \quad -G_{\zeta\zeta}C_N\\
-G_{\zeta\zeta}C_M & \quad G_{\zeta\zeta}
\end{pmatrix},
\ee
where $L$ is a characteristic length scale of the geometry, such that the displayed metric components, as well as chosen coordinates, are unitless. The proper size of the compactified direction is $d\hat{s}^2\equiv L^2R_{kk}^2e^{-2\Sigma}d\zeta^2$, where we have introduced the dimensionless Kaluza-Klein radius $R_{kk}$. To recover a light-like compactification we need to take a limit $R_{kk}\to 0$. Our formal dimensionless expansion parameter is this radius, but defining\footnote{One still has $\hbar=1$, so energy is measured in inverse time. $E=mc^2$ (or more precisely $KE=mv^2/2$) tells us that $mc=E/c$ has units of inverse length.} $R_{kk} \equiv (L m_{kk} c)^{-1}$ in terms of a Kaluza-Klein mass we can instead take the formal $c\to\infty$ limit. We will work in units with $L=1$ and identify the Kaluza-Klein mass $m_{kk}$ with the non-relativistic field theory mass $m$. The bulk proper Kaluza-Klein mass on the otherhand is $me^\Sigma$. We emphasize that this limit is simply a coordinate scaling limit: we
are taking the proper size of the compact direction to zero, while rescaling time such that the Kaluza-Klein mass remains finite.

Expanding the Kaluza-Klein gauge field as $C_M=c^2b_M+v_M$, and the asymptotic $d+2$ dimensional metric as
\begin{equation}
 \tilde{G}_{MN}=
\begin{pmatrix}
 -c^2N^2+N^KN_K & \quad c^2P_I+N_I\\
c^2P_J+N_J & \quad -c^2\frac{P_IP_J}{N^2}+G_{IJ}
\end{pmatrix},
\end{equation}
yields a line element, $d\hat{s}^2=\hat{G}_{XY}dx^Xdx^Y$, with pieces of $\mathcal{O}(c^2)$ and $\mathcal{O}(c^0)$, as well as vanishing negative powers of $c$. To be a non-singular consistent scaling limit of the $d+3$ dimensional geometry the $\mathcal{O}(c^2)$ pieces must vanish. Additionally, matching the $\mathcal{O}(c^0)$ components to those of the metric \ref{metricson} yields restrictions and identifications. Examining the $\mathcal{O}(c^2)$ term of the $dt^2$ piece we obtain the asymptotic restriction on the fields
\begin{equation}
 \label{gzz} me^\Sigma = \frac{b_t}{N}.
\end{equation}
This is the bulk implementation of the field theory constraint \ref{sigmadep}, which is the requirement that the chemical potential compensates the rest energy and allows the NR limit.
Combining this with the ``null'' $dtd\zeta$ and $dx^id\zeta$ pieces, and matching to the metric \ref{metricson}, we obtain the identifications
\begin{eqnarray}
 \label{phidef} \frac{e^{-\Phi}}{m}&\equiv&\frac{r^2N^2}{b_t},\\
\label{Bdef} B_i&\equiv&-\frac{b_i}{b_t},
\end{eqnarray}
where it is understood that this is a matching of the asymptotic $r\to0$ fields. The vanishing of the $\mathcal{O}(c^2)$ term of the $dtdx^i$ piece yields the restriction
\be
P_I=-N^2\frac{b_I}{b_t},
\ee
which, we recall from section \ref{sec:relpar}, encodes the requirement of the existence of a global time.
Matching the remaining metric components to \ref{metricson} we obtain the identifications
\begin{eqnarray}
\label{a0def} A_t&\equiv&v_t+\frac{b_tN^IN_I}{2N^2},\\
\label{aidef} A_i&\equiv&v_i+\frac{b_tN_i}{N^2}-\frac{b_iN^IN_I}{2N^2},\\
\label{metricdef} g_{ij}&\equiv&r^2\left(G_{ij}-\frac{b_iN_j}{b_t}-\frac{b_jN_i}{b_t}+\frac{b_ib_jN^IN_I}{b_t^2}\right).
\end{eqnarray}

It should be noted that the same partial gauge fixing which yielded the $d+3$ dimensional metric \ref{metricson} has been used to set $\hat{G}_{rr}=1/r^2$ and $\hat{G}_{r\zeta}=\hat{G}_{r\mu}=0$. In terms of the Kaluza-Klein fields this can be seen to yield
\begin{eqnarray}
G_{rr}=\frac{1}{r^2},\quad b_r=P_r=0,\quad N_r+\frac{e^{-2\Sigma}}{m^2}b_tv_r=0,\quad G_{ri}+\frac{e^{-2\Sigma}}{m^2}v_rb_i=0.
\end{eqnarray}
Extending the above definitions \ref{Bdef}, \ref{aidef}, and \ref{metricdef} to hold when an index is $r$ we see this partial gauge fixing gives $B_r=A_r=g_{ri}=0$.

Compared to the field theory non-relativistic limit \ref{RNRdict} the above identifications are equivalent, up to powers of $r$. While the fields $(\Phi,B_i,A_t,A_i,g_{ij})$ of metric \ref{metricson} are functions of only the field theory coordinates $t$ and $x_i$, the Kaluza-Klein fields $(\Sigma,P_I,N,N_I,b_M,v_M,G_{IJ})$ generically depend on the holographic radial direction as well\footnote{We are considering the $\zeta$ independent modes in each case corresponding to unbroken $U(1)$ invariance.}. The above identifications can be taken to tell us the asymptotic $r$ behavior of these fields. From \ref{Bdef} we see that $b_t$ and $b_i$ must have the same asymptotic behavior, which combined with \ref{metricdef} gives the leading asymptotic behavior of $G_{ij}$ and $N_i$ as $r^{-2}$. From \ref{phidef} we see that $N^2/b_t$ goes as $r^{-2}$, while \ref{a0def} and \ref{aidef} determine $v_M$ to be asymptotically independent of $r$.

Further determination requires assumptions on the behavior of $\Sigma$. For the asymptotic form $e^{-\Sigma}\equiv e^{-\sigma(t,\vec{x})}/r^{\delta}$ and using \ref{gzz} we obtain the asymptotic behaviors $N\sim r^{\delta-2}$ and $b_M\sim r^{2\delta-2}$. Note that for $\delta=1$ the lapse goes as $N\sim r^{-1}$ and the metric is asymptotically AdS. One can extend the symmetries to include the non-relativistic conformal transformations of \ref{conf} by considering radial diffeomorphisms, as in \cite{Son:2008ye}, which in fact fix $\delta=1$. These transformations will be more fully explored in the next section.

The Kaluza-Klein viewpoint illuminates the factor of $me^\sigma$ arising in the non-relativistic Lagrangian density derived by the $c\to\infty$ limit of the relativistic field theory in section \ref{sec:relpar}. Upon dimensional reduction the volume density of the higher dimensional theory yields the lower dimensional volume density, as well as a factor related to the proper Kaluza-Klein mass. In our case this gives an overall factor of $\sqrt{G_{\zeta\zeta}}=e^{-\Sigma}/mc$ causing the non-relativistic Lagrangian to be exactly that of \ref{NRaction}, even for spacetime dependent $m$ and $\Sigma$.

\subsubsection{Einstein-Maxwell vector khronon}\label{sec:EMvk}\label{sec:EMFT}
To the point of excess, we now present a more general derivation of a holographic map relating bulk and NR QFT fields. The motivation follows from the previous sections: it was seen that GR on a $d+3$ dimensional manifold can capture the generic symmetries of a $d+1$ dimensional NR QFT by taking a particular compactification and scaling limit. This specific duality is overly restrictive; despite containing fields that obey NR electro-diffeomorphism invariance most of the properties are simply inherited from the relativistic derivation.

We start with the Kaluza-Klein reduced field content of section \ref{sec:kkk}, a graviton and a Maxwell field (the scalar will not play a role here), and show that the NR limit can be taken directly in Einstein-Maxwell theory.
Previously, spatial compactification and a scaling limit gave a light-like compactification of $d+3$ dimensional general relativity. We will now start with the $d+2$ dimensional field content of the Kaluza-Klein theory, that is the Einstein-Maxwell system, and take a true $d+2$ dimensional\footnote{This is to contrast with the scaling limit of the previous section. There, after the $c\to\infty$ limit, we still had a finite $d+3$ dimensional spacetime metric. If the Einstein-Maxwell fields were recombined back into a higher dimensional spacetime metric it would contain non-sensible $\mathcal{O}(c^2)$ pieces. In this section we simply take the fields that survived in this limit, but do not take the specific form of the Kaluza-Klein action with a scalar field dependent gauge kinetic term.} non-relativistic $c\to\infty$ limit.

The relativistic diffeomorphism generators are expanded as $\xi^M=(f-\alpha/c^2,\xi^I)$, under which the $d+2$ dimensional metric transforms as
\begin{equation}
 \delta \tilde{G}_{MN} = \xi^P\d_P\tilde{G}_{MN}+\tilde{G}_{MP}\d_N\xi^P+\tilde{G}_{NP}\d_M\xi^P.
\end{equation}
For the consistent expansion
\begin{equation}\tilde{G}_{MN}\equiv
 \begin{pmatrix}
  -c^2N^2-2N^2A+N^KN_K & \quad c^2P_I+N_I\\
c^2P_J+N_J & \quad -c^2\frac{P_IP_J}{N^2}+G_{IJ}
 \end{pmatrix},
\end{equation}
under the diffeomorphism transformations in the $c\to\infty$ limit, the metric fields transform as:
\begin{eqnarray}
\nn\delta N&=&\xi^K\d_KN+f\dot{N}+\dot{f}N-\frac{P_K}{N}\dot{\xi}^K,\\
\nn\delta A &=&\xi^K\d_KA+f\dot{A}-(\dot{\alpha}-N^I\d_I\alpha)\left(1+\frac{N^KP_K}{N^2}\right)+2\frac{AP_K}{N^2}\dot{\xi}^K-2AN^K\d_Kf,\\
\nn\delta N_I&=& \xi^K\d_KN_I+N_K\d_I\xi^K+f\dot{N}_I+\dot{f}N_I+G_{IK}\dot{\xi}^K+\d_If\left(N^KN_K-2N^2A\right)+N^2\d_I\alpha-\dot{\alpha}P_I,\\
\nn\delta G_{IJ}&=& \xi^K\d_KG_{IJ}+G_{IK}\d_J\xi^K+G_{JK}\d_I\xi^K+f\dot{G}_{IJ}+N_I\d_Jf+N_J\d_If-P_I\d_J\alpha-P_J\d_I\alpha,\\
\label{gtrafo}\delta P_I&=&\xi^K\d_KP_I+P_K\d_I\xi^K+f\dot{P}_I+\dot{f}P_I-\frac{P_IP_K}{N^2}\dot{\xi}^K-N^2\d_If.
\end{eqnarray}

The relativistic Maxwell gauge field can be expanded as $C_M\equiv c^2b_M+v_M$. It transforms under the action of the gauge generator $\Lambda\equiv c^2\beta+\lambda$ and the relativistic diffeomorphisms $\xi^M$ as
\begin{equation}
 \delta C_M=\xi^N\d_NC_M+C_N\d_M\xi^N-\d_M\Lambda.
\end{equation}
Taking the $c\to\infty$ limit gives the transformations for the gauge fields:
\begin{eqnarray}
 \nn\delta b_t&=&\xi^K\d_Kb_t+f\dot{b}_t+\dot{f}b_t+b_K\dot{\xi}^K-\dot{\beta},\\
\nn\delta b_I&=&\xi^K\d_Kb_I+b_K\d_I\xi^K+f\dot{b}_I+b_t\d_If-\d_I\beta,\\
\nn\delta v_t&=&\xi^K\d_Kv_t+f\dot{v}_t+\dot{f}v_t+v_K\dot{\xi}^K-\dot{\lambda}-b_t\dot{\alpha},\\
\label{vectrafo}\delta v_I&=&\xi^K\d_Kv_I+v_K\d_I\xi^K+f\dot{v}_I+v_t\d_If-\d_I\lambda-b_t\d_I\alpha.
\end{eqnarray}

Lastly, we consider a complex scalar $\Psi$ charged under the gauge field. It has the relativistic transformation
\begin{equation}
 \delta \Psi =\xi^M\d_M\Psi-\imath \Lambda \Psi.
\end{equation}
Expanding the field as $\Psi\equiv \rho e^{-\imath\eta}$ for $\eta\equiv c^2\phi +\chi$, in the $c\to\infty$ limit, the real magnitude and phases transform as
\begin{eqnarray}
\nn \delta \rho&=&\xi^K\d_K\rho+f\dot{\rho},\\
\nn \delta \phi&=&\xi^K\d_K\phi +f\dot{\phi}+\beta,\\
\delta \chi&=&\xi^K\d_K\chi+f\dot{\chi}+\lambda-\dot{\phi}\alpha.
\end{eqnarray}
In the background we are considering we will work with $\Psi=0$ in the end, so this particular form of the matter fields is not essential. What we need is that some charged matter exists in the bulk, so that constant $A_t$ can not be simply gauged away. In the Kaluza-Klein example of the previous subsection the role of the charged matter was played by the massive Kaluza-Klein gravitons.

This procedure has given us a consistent set of fields that transform sensibly in the $c\to\infty$ non-relativistic limit. To go further, for example to construct an action and determine which fields have non-trivial dynamics, we will make some simplifying restrictions. Most importantly, we require the theory to have a global time. As discussed in section \ref{sec:NRQFTintro} this is necessary to have a causal non-relativistic theory. It can be implemented by constructing a spacetime foliation whose leaves contain events that happen at the same global time. Parallel to the previous discussion, this can be achieved by considering a scalar field whose level sets define the foliation leaves. The shortcomings of this scalar khronon formalism, enumerated in section \ref{sec:scalar}, requires a different approach in the pursuit of a bulk theory.

These problems will be circumvented by considering $\phi$ to be the gauged phase of a charged field, but we will not define a global time via its level sets. Instead, given the expectation value $\phi=t$ we will set this phase to zero by performing the gauge transformation $\beta=-t$, which will turn on a constant time component of the gauge field, $b_t$. Thus the vector $b_M$ acts as a ``khronon'' and determines the foliation by a global time: when in adapted coordinates it has only a temporal component. Once the expectation value of $\phi$ has been gauged away, in order to preserve $\phi=0$, we can no longer perform the ``large'' gauge transformations $\beta$. We still have time reparametrizations as performing a spatially independent $f(t)$ maintains $b_I=0$, that is, it keeps us with a physical global time.

\subsection{NR holography}
\subsubsection{Holographic map}\label{sec:hmap}
By examining the above transformations of bulk fields, we can determine combinations which asymptotically transform as \ref{QFTtrafo}. Firstly, for $\beta=0$, the two combinations
\be -\frac{b_i}{b_t},\quad \frac{P_i}{N^2}, \ee
both transform as the non-relativistic field $B_i$, with which we identify them. This relation between the metric field $P_I$ and the gauge field $b_I$, as discussed in section \ref{sec:relpar}, is required for the existence of a global time. It is then seen that both $N$ and $b_t$ transform like $e^{-\Phi}$, and in generality we asymptotically identify
\be \label{hphidef}e^{-\Phi}\equiv r^{\gamma(\delta_\Phi+1)}N\left(\frac{N}{b_t}\right)^{\delta_\Phi},\ee
where the factor $r^{\gamma(\delta_\Phi+1)}$ is required to strip off the asymptotic radial behavior of the bulk fields, and $\delta_\Phi$ is an arbitrary power. This parametrization assumes that asymptoticaly $b_t\sim r^0$, which is natural for the vector khronon, and that therefore $N\sim 1/r^\gamma$. We will shortly find restrictions on $\delta_\Phi$ and $\gamma$ due to the conformal dimensions of the NR fields. Lastly, it can be seen that the combinations
\begin{eqnarray}
 \nn A_t&\equiv& v_t+\left(\frac{b_t}{N}\right)^{\frac{2}{\gamma}-1}\left(\frac{N^IN_I}{2N}-NA\right),\\
 \nn A_i&\equiv& v_i+\left(\frac{b_t}{N}\right)^{\frac{2}{\gamma}-1}\left(\frac{N_i}{N}-\frac{b_i}{b_t}\left(\frac{N^IN_I}{2N}-NA\right)\right),\\
 \label{holomap} g_{ij}&\equiv& r^2\hat{g}_{ij}=r^2\left(G_{ij}-\frac{b_iN_j}{b_t}-\frac{b_jN_i}{b_t}+2\frac{b_ib_jN}{b_t^2}\left(\frac{N^IN_I}{2N}-NA\right)\right),
\end{eqnarray}
asymptotically transform under $f$, $\xi^i$, and $\lambda$ as the field theory gauge fields and metric if we make the identification
\be
m\equiv r^{\gamma(\delta_\Phi+1)-2}\left(\frac{b_t}{N}\right)^{\frac{2}{\gamma}-\delta_\Phi-1}.
\ee
This requirement comes from examining the transformation of $A_i$, and equating the coefficient of $\hat{g}_{ij}\dot{\xi}^j$ with the bulk fields corresponding to $me^\Phi$, to reproduce \ref{QFTtrafo}.

\subsubsection{Subleading temporal diffeomorphisms}\label{sec:alpha}
We now come to the overdue discussion of the role of the subleading temporal diffeomorphism $\alpha$. The field theory quantities are not affected by this transformation, as seen in \ref{QFTtrafo}. There are two different scenarios for the role of $\alpha$ in the bulk; both of them have an interesting holographic interpretation and lead to physically distinct pictures. One option is that the bulk action is not invariant under $\alpha$ transformations, we therefore never perform this transformation in the bulk. This is a consistent truncation of the $c\to \infty$ expansion, and also allows us to set $A=0$; we need not consider the subleading expansion of the lapse $N$. The above then gives a well defined dictionary between bulk and field theory quantities, parametrized by the two constants $\gamma$ and $\delta_\Phi$. The fields defined in \ref{holomap} are then just a part of the boundary sources; there are additional gauge invariant bulk fields, such as e.g. $N_r$, and hence also additional field theory sources.

Alternatively, the subleading temporal diffeomorphism $\alpha$ can be a gauge invariance of the bulk theory. That is, it can be interpreted as a redundancy of the bulk description, and therefore should not effect the field theory data. The fields defined in \ref{holomap} are only invariant under $\alpha$ for $\gamma =1$, or equivalently $N\sim 1/r$. Appearing mysterious in the Kaluza-Klein derivation of section \ref{sec:kkk}, this justifies the combinations of bulk fields that give the field theory ones. As that bulk theory contains the full diffeomorphism invariance of GR, the only physical boundary fields are those that are invariant under the bulk redundancy $\alpha$, and therefore the ones appearing in \ref{holomap} with $\gamma=1$. This also elucidates the appearance of the subleading temporal diffeomorphism in the NR QFT work of \cite{Son:2005rv} and the generally covariant Ho\v rava-Lifshitz theory of \cite{Horava:2010zj,daSilva:2010bm}. As they inherently consider uncharged fields they do not have the explicit gauge field $v_\mu$. From \ref{vectrafo}, to consistently consider the transformation $\alpha$, but to maintain $v_\mu=0$, one must implicitly perform a gauge transformation $\lambda$. The $\alpha$ variant piece $v_I$ of the invariant $A_I$, defined above, was held fixed. Thus the redundancy $\alpha$ was made physical by linking it
to the transformation $\lambda$, which is a global symmetry of the field theory.

\subsubsection{NR  scale and conformal invariance}\label{sec:Hconf}
We additionally would like to be able to describe NR QFTs that have the NR conformal symmetry of \ref{zne2}. As with traditional holography, this transformation is captured by symmetries of the bulk theory. Unlike the usual AdS/CFT correspondence, these symmetries are not strict isometries of the spacetime geometry, but instead manifest as transformations acting on the above combinations identified as field theory quantities. As in traditional holography and \cite{Son:2008ye}, the conformal structure of the field theory is captured by radial diffeomorphisms in the bulk. Under $\xi^r=-\omega(t,\vec{x}) r$ the field theory data transform as
\begin{eqnarray}
\nn \delta e^{-\Phi}&\supset& r^{\gamma(\delta_\Phi+1)}\xi^r\d_r \left(N\left(\frac{N}{b_t}\right)^{\delta_\Phi}\right)=\gamma(\delta_\Phi+1)\omega e^{-\Phi},\\
\nn \delta g_{ij} &\supset& r^2\xi^r\d_r \left(\hat{g}_{ij}\right)=2\omega\hat{g}_{ij}r^2=2\omega g_{ij},\\
\delta m &\supset& r^{\gamma(\delta_\Phi+1)-2}\xi^r\d_r\left(\frac{b_t}{N}\right)^{\frac{2}{\gamma}-\delta_\Phi-1}=\left(\gamma(\delta_\Phi+1)-2\right)\omega m,
\end{eqnarray}
which agrees with the field theory conformal transformation \ref{zne2} for $\delta_\Phi=z/\gamma-1$. The most interesting case for us, with the bulk being AdS (that is $N\sim1/r$) and a mass invariant under scale transformations, corresponds to
\be
\gamma=1, \quad z=2, \quad \delta_\Phi=1.
\ee

For NR general covariance the bulk transformations that preserve the trivial asymptotic background $\Phi=B_I=A_t=A_I=0$ and $G_{IJ}=\delta_{IJ}/r^2$, should agree with the field theory symmetries. The first case of a scale transformation starts with the temporal rescaling $f=\kappa t$. To maintain $\Phi=0$, from above, we see that we require the radial rescaling $\xi^r=\kappa r/z$. To maintain $G_{ij}$ the spatial rescaling $\xi^i=\kappa x^i/z$ must be performed, in agreement with a dynamical critical exponent of $z$.  Lastly, $B_I$, $G_{rI}$, and $A_M=0$ are automatically maintained under these scale transformations. In complete parallel to the field theory discussion in section \ref{sec:FTconf}, $m$ changes for $z\ne 2$ and so in this case the symmetry is only spurionic. Although these bulk combinations have the same isometries as the field theory quantities with which we identify them, the bulk fields themselves may not be invariant. Under the scale transformations we see that generically
\begin{equation}
 \delta b_t = \kappa b_t,\quad \delta N=\kappa \left(1-\frac{\gamma}{z}\right)N.
\end{equation}
As we will discuss further below, this non-invariance of $N$ can be interpreted as evidence for hyperscaling violation of the theory. On the other hand, for the bulk action of probe fields, we expect bulk fields to enter only in the combinations identified above. Here factors of $b_t/N$ act like the $\sigma$ field of previous sections, adding it to the action can change the dimension of the probe fields.

For the special case of $z=2$ there is an additional transformation of the bulk fields preserving the trivial background. This ``special conformal'' transformation involves the time reparametrization $f=Ct^2$. To preserve $\Phi=0$, from above, we must also perform the radial redefinition $\xi^r=Ctr$. Preservation of the trivial metric then requires $\xi^i=Ctx^i$. Lastly, maintaining the form of the bulk fields that correspond to the trivial gauge configuration requires the gauge transformation $\lambda=C(\vec{x}^2+r^2)/2$. As with the scale transformation, not all bulk fields are invariant under this special conformal transformation. In addition to $N$ and $b_t$, and the issues discussed above, the shift vector is not invariant under the time dependent $\xi^I$, but transforms as
\be \delta N_I=\frac{C x^I}{r^2}. \ee
These fields should correspond to gauge invariant operators in the field theory, and thus it appears that NR conformal invariance is generically untenable. It can be recovered for the special case of bulk invariance under the subleading temporal diffeomorphism $\alpha$. This transformation allows the shift $N_I$ to be held to zero, as well as the maintenance of $A=0$ for the subleading term of the lapse. As shown above, $\alpha$ invariance restricts $N\sim 1/r$, that is, the bulk background is that of AdS. We are therefore able to realize the NR scale and conformal isometries of section \ref{sec:FTconf}.

\subsubsection{Bulk action}\label{sec:bulkact}
Consider, initially, bulk theories without the $\alpha$ transformation. This also allows us to consistently set $A=0$; we do not need to consider the subleading piece of the lapse $N$ in the $c$ expansion. As previously discussed, by working in a global time we can also maintain $b_I=P_I=0$. This gives us the following consistent field content: the metric is decomposed in the ADM variables $N$, $N_I$, and $G_{IJ}$ adapted to the preferred foliation; the gauge vector behaves as the non-relativistic decomposition $v_t$ and $v_I$ with respect to the global time. The background ``large'' gauge field $b_t$ determines the foliation by a global time, and should be considered a parameter that must be tuned to yield a NR holographic duality, much like the cosmological constant in traditional holography. The non-redundant transformations are spatial diffeomorphisms $\xi^I$, temporal reparametrization $f(t)$, and the $U(1)$
transformation $\lambda$. This is exactly the field content and symmetries of Ho\v rava gravity coupled to non-relativistic electromagnetic fields: our proposal for a holographic dual to a generic NR QFT obeying the symmetries \ref{QFTtrafo} is this non-relativistic gravity theory, on a background spacetime with a non-zero $b_t$. The bulk action will therefore be determined by the couplings $\tilde{\lambda}$ and $\alpha$ of Ho\v rava gravity, as well as those introduced with non-relativistic electromagnetic fields\footnote{Besides the gauge coupling we have one more parameter which gives the relative size of the $\vec{E}^2$ and $\vec{B}^2$ terms in the action; the effective speed of light for the electromagnetic field is a free parameter and not fixed to be $c$, very much like in the textbook treatment of electromagnetism in matter.}.
To go further, we note that the covariant Ho\v rava-Lifshitz theory of \cite{Horava:2010zj,daSilva:2010bm}, coupled to electromagnetic fields, is a bulk theory with $\alpha$ invariance and the same fields and symmetries as above. It therefore is capable of holographically describing Schr\"odinger invariant NR CFTs.

\subsubsection{Background solutions and correlation functions}\label{sec:corrfunc}
A class of simple solutions to Ho\v rava gravity is motivated by the scalar khronon formulation of section \ref{sec:scalar}. There it was argued that for parametrically small $\tilde{\lambda}$ and $\alpha$ solutions of Einstein gravity descend to solutions of Ho\v rava gravity. This was justified as the scalar khronon $\chi$ had a stress tensor of order $(\tilde{\lambda},\alpha)$ and therefore acted as a probe and did not backreact. Here we simply use it as motivation: using Einstein solutions as a leading order ansatz we then look for solutions to the action \ref{Haction} to leading order in these small parameters. For the negative cosmological constant $\Lambda=-d(d+1)/2$ and no bulk gauge fields we obtain the leading fields:
\begin{eqnarray}
 \nn G_{ij}&=&\frac{\delta_{ij}}{r^2},\\
 \nn G_{rr}&=&\frac{1}{r^2}+\alpha \frac{2d+1}{d(d+1)r^2},\\
 N&=&\frac{1}{r}-\alpha\frac{\log[r]}{r}.
\end{eqnarray}
In this formulation the new degree of freedom in Ho\v rava gravity is captured by the shift component $N_r$, which is zero in GR. Linearizing its equation of motion we obtain the near boundary behavior
\begin{equation}
N_r\sim r^{\left(\frac{d-3}{2}\pm\frac{1}{2}\sqrt{(d+1)^2+\frac{4d\alpha}{\tilde{\lambda}}}\right)}.
\end{equation}
This structure implies a Breitenlohner-Freedman \cite{Breitenlohner:1982jf} like bound on the values these parameters can take to avoid rapidly oscillating (and hence presumably unstable) solutions for $N_r$. In order for the exponents to be real, we need
\begin{equation}
\frac{\alpha}{\tilde{\lambda}} \geq - \frac{(d+1)^2}{4 d}.
\end{equation}

As discussed in section \ref{sec:hmap} the asymptotic behavior of the lapse $N$ function is captured by the exponent $\gamma$ and sets the form of the holgraphic map. To zeroth order in $\alpha$, the above metric is that of AdS and the scaling is $N\sim 1/r$. The appearance of the $\log$ correction to $N$ is a signal that this scaling is modified asymptotically and the corrected exponent is $1+\alpha$. Indeed, moving away from the ``probe'' limit of small $\tilde{\lambda}$ and $\alpha$, and making the ansatz $N_r=0$, one finds\footnote{This solution was first brought to our attention in private discussion by Charles Melby-Thompson.} the solution
\begin{eqnarray}
 \nn \Lambda &=& \frac{-(d(d+1)+\alpha(2d^2-1)-\alpha^2d(d-1)}{2(1-\alpha)^2},\\
 G_{IJ}&=&\frac{\delta_{IJ}}{r^2},\quad N=r^{\frac{1}{\alpha-1}}.
\end{eqnarray}
Most interestingly the radial behavior of the lapse $N$, for $\alpha<1$, seems capable of reproducing an arbitrary $\gamma$. In the holographic context this background solution undoubtedly deserves further study.

To calculate correlation functions one needs to examine the on-shell action of the dual bulk fields. For a bulk scalar with $z=2$, to motivate the form of the action, we will take the non-relativistic action \ref{NRaction} written in terms of the Ho\v rava fields. This gives the following bulk action for a charged scalar
\begin{equation}
\label{scalaraction}
 S_\Psi=\int dt dr d^dx\sqrt{G}\frac{N^2}{b_t}\left[\left(\frac{\imath b_t}{2N^2}\Psi^\dagger\left(\mathcal{D}_t-N^J\mathcal{D}_J\right)\Psi+h.c.\right)-\frac{G^{IJ}}{2m}\mathcal{D}_I\Psi^\dagger\mathcal{D}_J\Psi-\frac{M^2}{2m}\Psi^\dagger\Psi\right],
\end{equation}
where the metric and gauge covariant derivatives are given by $\mathcal{D}_t=\d_t-\imath v_0$, $\mathcal{D}_I=\nabla_I-\imath v_I$, and we have included the non-relativistic bulk mass term $M$. The combination of temporal and spatial derivatives in the kinetic term is expected for invariance under foliation preserving diffeomorphisms. To zeroth order in $\alpha$ we have found AdS as a background solution. On this background the scalar action becomes
\begin{equation}
 S_\Psi=\int dt dr d^dx \frac{1}{b_t r^{d+3}}\left[\imath b_t r^2\Psi^\dagger\d_t\Psi-\frac{r^2}{2m}\d_I\Psi^\dagger\d_I\Psi-\frac{M^2}{2m}\Psi^\dagger\Psi\right],
\end{equation}
where we have assumed $m$ and $b_t$ are constant. This agrees with the action of \cite{Son:2008ye} up to an overall constant. We therefore can copy their calculation of the correlation function of the field theory operator dual to this scalar. In momentum space this gives
\begin{equation}
 \langle\mathcal{O}\mathcal{O}\rangle\sim\left(\vec{k}^2-2mb_t\omega\right)^{2\nu},
\end{equation}
where \begin{equation}
       \nn \nu=\sqrt{\frac{(d+2)^2}{4}+M^2}.
      \end{equation}
Upon Fourier transforming to real space this gives the restrictive form dictated by Galilean and scale symmetry \cite{Henkel:1993sg,henkel,Nishida:2010tm}, providing a quantitative check of the duality. Comparison to \cite{Son:2008ye,Balasubramanian:2008dm} shows that the constant $mb_t$ plays the role
of the charge or particle number of the operator $\mathcal{O}$.

This form of the action can  also be motivated as a derivative expansion. At zero derivatives we have simply the mass term. At one derivative, using the NR fields, we can construct the terms
\begin{equation}
 \Psi^\dagger b^t\d_t\Psi,\quad\Psi^\dagger N^I\d_I\Psi,\quad \Psi^\dagger b^I\d_I\Psi,\quad \Psi^\dagger P^I\d_I\Psi.
\end{equation}
For Ho\v rava gravity the last two terms are absent, while the first two are taken in the combination that is invariant under foliation preserving diffeomorphisms. At two derivatives the leading term is simply the canonical spatial gradient squared term. Other bulk probe actions are possible, given only the symmetry restrictions of Ho\v rava gravity. In particular, the Lagrangian can be multiplied by the overall factor $(b_t/N)^\Theta$. The effect of this factor is to shift the dimension of the operator coupled to the bulk field, mimicking the $\sigma$ field of section \ref{sec:relpar}. It can also be understood to represent hyperscaling violation, as the dimension of the operator is changed by replacing $d\to d-\Theta$, modifying the effective number of spatial dimensions of the theory. The scalar action with $\Theta=1$ in many ways appears to be the most natural. In that case no inverse powers of $b_t$ appear in the action and the potential simply has an overall prefactor of $N$ as part of the usual
measure. This is exactly the scalar action one would have written down in Ho\v rava gravity without the extra Maxwell field.

\section{Discussion: String theory embeddings}\label{sec:discussion}
We have argued that, based on its symmetry structure, Ho\v rava gravity is the natural holographic dual of a generic NR QFT. To make sure our ideas are correct, it would of course be nice to confirm that our construction can be consistently embedded into string theory. This embedding is facilitated by our observation that we can derive NR systems quite generically as a $c \rightarrow \infty$ limit of a relativistic theory by setting the chemical potential equal to the rest energy of the lightest charged particle. All we need to do in order to give string theory embeddings of our scenario is to find relativistic examples of holographically dual pairs where the field theory side has a global $U(1)$ symmetry with massive charged particles.

One such example was in fact already presented in section \ref{sec:kkk}. We can start with the known duality between AdS$_5$ $\times$ $S^5$ in type IIB string theory and ${\cal N}=4$ super Yang-Mill gauge theory and
compactify the latter on a circle with periodic boundary conditions. In this case the resulting $2+1$ dimensional relativistic field theory has a new global $U(1)$ symmetry associated with shifts around the compact circle.
The charged particles are the momentum modes in the internal direction and they naturally have a mass equal
to the inverse circle radius. The non-relativistic limit in this theory introduces a chemical potential
for this $U(1)$ particle number equal to the rest mass of the KK particles and then takes the $c \rightarrow \infty$ limit. This is exactly what we did in section \ref{sec:kkk} where we showed that in this limit the
circle becomes light-like. In this KK example, a massless scalar in the relativistic geometry (for example the IIB dilaton) in terms of the Ho\v rava gravity variables is exactly given by the action \ref{scalaraction}. Here the at first sight unnatural $N/b_t$ prefactor we introduced in the action in order to avoid hyperscaling violating comes from the higher dimensional origin; it is exactly the $\sqrt{G_{\zeta \zeta}}$ prefactor in the bulk action we alluded to at the end of section \ref{sec:kkk}.

While many examples of holographic dualities in the presence of finite chemical potential are understood by now, the task of finding additional examples where the charge carriers are massive so that we can implement the NR limit advertised here is somewhat more non-trivial. One example is ABJM theory which allows a supersymmetric preserving mass term and a non-relativistic limit \cite{Nakayama:2009cz,Lee:2009mm}. Gravitational solutions of M-theory matching the global symmetries of this NR CFT were studied in detail in \cite{Jeong:2009aa}. There it was found that the prospective gravity dual did not have the same amount of supersymmetry as the NR ABJM field theory. This leads one to question the role of supersymmetry in non-relativistic holograpy. Although it is crucial in traditional AdS/CFT, often providing stability to the best known examples, it may not be as important for NR physics.

Another large class of examples of holographically dual pairs with a finite density of massive charge carriers is based on probe branes \cite{Karch:2002sh} which were first studied at finite chemical potential in \cite{Kobayashi:2006sb}. In this situation the thermodynamics and the spectrum of hydrodynamic modes was recently analyzed exactly in the NR limit advertised here \cite{Karch:2007br,Ammon:2012je}. While in those papers the results were not phrased in the language of Horava gravity, the findings especially of the latter are completely consistent with the picture we developed here. In the scaling limit the probe brane system is found to be governed by a NR CFT with $z=2$ and $\Theta=1$ for two physically quite distinct probe systems
(with $d=3$ and $d=2$ spatial dimensions respectively).

Obviously many interesting questions still remain. Maybe most interestingly will be to study the thermodynamic properties of NR systems that do not simply follow by a scaling limit from a relativistic theory. This can safely be done in the large $N_c$ limit by studying black hole solutions to Ho\v rava-Maxwell theory with parameters $\tilde{\lambda}$ and $\alpha$ away from the probe limit. Of course even more interesting will be to tackle quantum Ho\v rava theory and move away from large $N_c$.

\section*{Acknowledgements}
We'd like to thank Carlos Hoyos, Charles Melby-Thompson, Sergej Moroz and Edward Witten for helpful discussions. AK also would like to thank the members of the Stanford Institute of Theoretical Physics, in particular Shamit Kachru, Steve Shenker, Eva Silverstein, and Lenny Susskind for important feedback on a presentation of a preliminary version of this work. Last but not least, special thanks to Dam Son for his patient explanations of his numerous contributions to this subject and for many very valuable discussions.
This work was supported in part by the U.S. Department of Energy under Grant No.~DE-FG02-96ER40956.

\bibliographystyle{JHEP}
\bibliography{NRHolo}

\providecommand{\href}[2]{#2}\begingroup\raggedright\begin{thebibliography}{10}

\bibitem{Son:2005rv}
D.~Son and M.~Wingate, {\it {General coordinate invariance and conformal
  invariance in nonrelativistic physics: Unitary Fermi gas}},  {\em Annals
  Phys.} {\bf 321} (2006) 197--224,
  [\href{http://xxx.lanl.gov/abs/cond-mat/0509786}{{\tt cond-mat/0509786}}].

\bibitem{Hoyos:2011ez}
C.~Hoyos and D.~T. Son, {\it {Hall Viscosity and Electromagnetic Response}},
  {\em Phys.Rev.Lett.} {\bf 108} (2012) 066805,
  [\href{http://xxx.lanl.gov/abs/1109.2651}{{\tt arXiv:1109.2651}}].

\bibitem{Maldacena:1997re}
J.~M. Maldacena, {\it {The Large N limit of superconformal field theories and
  supergravity}},  {\em Adv.Theor.Math.Phys.} {\bf 2} (1998) 231--252,
  [\href{http://xxx.lanl.gov/abs/hep-th/9711200}{{\tt hep-th/9711200}}].

\bibitem{Witten:1998qj}
E.~Witten, {\it {Anti-de Sitter space and holography}},  {\em
  Adv.Theor.Math.Phys.} {\bf 2} (1998) 253--291,
  [\href{http://xxx.lanl.gov/abs/hep-th/9802150}{{\tt hep-th/9802150}}].

\bibitem{Gubser:1998bc}
S.~Gubser, I.~R. Klebanov, and A.~M. Polyakov, {\it {Gauge theory correlators
  from noncritical string theory}},  {\em Phys.Lett.} {\bf B428} (1998)
  105--114, [\href{http://xxx.lanl.gov/abs/hep-th/9802109}{{\tt
  hep-th/9802109}}].

\bibitem{Son:2008ye}
D.~Son, {\it {Toward an AdS/cold atoms correspondence: A Geometric realization
  of the Schrodinger symmetry}},  {\em Phys.Rev.} {\bf D78} (2008) 046003,
  [\href{http://xxx.lanl.gov/abs/0804.3972}{{\tt arXiv:0804.3972}}].

\bibitem{Balasubramanian:2008dm}
K.~Balasubramanian and J.~McGreevy, {\it {Gravity duals for non-relativistic
  CFTs}},  {\em Phys.Rev.Lett.} {\bf 101} (2008) 061601,
  [\href{http://xxx.lanl.gov/abs/0804.4053}{{\tt arXiv:0804.4053}}].

\bibitem{Horava:2010zj}
P.~Horava and C.~M. Melby-Thompson, {\it {General Covariance in Quantum Gravity
  at a Lifshitz Point}},  {\em Phys.Rev.} {\bf D82} (2010) 064027,
  [\href{http://xxx.lanl.gov/abs/1007.2410}{{\tt arXiv:1007.2410}}].

\bibitem{Horava:2009uw}
P.~Horava, {\it {Quantum Gravity at a Lifshitz Point}},  {\em Phys.Rev.} {\bf
  D79} (2009) 084008, [\href{http://xxx.lanl.gov/abs/0901.3775}{{\tt
  arXiv:0901.3775}}].

\bibitem{daSilva:2010bm}
A.~M. da~Silva, {\it {An Alternative Approach for General Covariant
  Horava-Lifshitz Gravity and Matter Coupling}},  {\em Class.Quant.Grav.} {\bf
  28} (2011) 055011, [\href{http://xxx.lanl.gov/abs/1009.4885}{{\tt
  arXiv:1009.4885}}].

\bibitem{Greenwald:2010fp}
J.~Greenwald, V.~Satheeshkumar, and A.~Wang, {\it {Black holes, compact objects
  and solar system tests in non-relativistic general covariant theory of
  gravity}},  {\em JCAP} {\bf 1012} (2010) 007.

\bibitem{Fefferman:1985}
C.~Fefferman and C.~R. Graham, {\it Conformal invariants},  {\em Ast\'erisque,
  Numero Hors Serie} {\bf 95} (1985).

\bibitem{Sotiriou:2009gy}
T.~P. Sotiriou, M.~Visser, and S.~Weinfurtner, {\it {Phenomenologically viable
  Lorentz-violating quantum gravity}},  {\em Phys.Rev.Lett.} {\bf 102} (2009)
  251601, [\href{http://xxx.lanl.gov/abs/0904.4464}{{\tt arXiv:0904.4464}}].

\bibitem{Horava:2008ih}
P.~Horava, {\it {Membranes at Quantum Criticality}},  {\em JHEP} {\bf 0903}
  (2009) 020, [\href{http://xxx.lanl.gov/abs/0812.4287}{{\tt
  arXiv:0812.4287}}].

\bibitem{Griffin:2011xs}
T.~Griffin, P.~Horava, and C.~M. Melby-Thompson, {\it {Conformal Lifshitz
  Gravity from Holography}},  \href{http://xxx.lanl.gov/abs/1112.5660}{{\tt
  arXiv:1112.5660}}.

\bibitem{Berglund:2012fk}
P.~Berglund, J.~Bhattacharyya, and D.~Mattingly, {\it {Thermodynamics of
  universal horizons in Einstein-aether theory}},
  \href{http://xxx.lanl.gov/abs/1210.4940}{{\tt arXiv:1210.4940}}.

\bibitem{short}
S.~Janiszewski and A.~Karch {\em to be published} (2012).

\bibitem{Greiter:1989qb}
M.~Greiter, F.~Wilczek, and E.~Witten, {\it {Hydrodynamic Relations in
  Superconductivity}},  {\em Mod.Phys.Lett.} {\bf B3} (1989) 903.

\bibitem{Blas:2009yd}
D.~Blas, O.~Pujolas, and S.~Sibiryakov, {\it {On the Extra Mode and
  Inconsistency of Horava Gravity}},  {\em JHEP} {\bf 0910} (2009) 029,
  [\href{http://xxx.lanl.gov/abs/0906.3046}{{\tt arXiv:0906.3046}}].

\bibitem{Blas:2010hb}
D.~Blas, O.~Pujolas, and S.~Sibiryakov, {\it {Models of non-relativistic
  quantum gravity: The Good, the bad and the healthy}},  {\em JHEP} {\bf 1104}
  (2011) 018, [\href{http://xxx.lanl.gov/abs/1007.3503}{{\tt
  arXiv:1007.3503}}].

\bibitem{Germani:2009yt}
C.~Germani, A.~Kehagias, and K.~Sfetsos, {\it {Relativistic Quantum Gravity at
  a Lifshitz Point}},  {\em JHEP} {\bf 0909} (2009) 060,
  [\href{http://xxx.lanl.gov/abs/0906.1201}{{\tt arXiv:0906.1201}}].

\bibitem{Kiritsis:2012ta}
E.~Kiritsis, {\it {Lorentz violation, Gravity, Dissipation and Holography}},
  \href{http://xxx.lanl.gov/abs/1207.2325}{{\tt arXiv:1207.2325}}.

\bibitem{ArkaniHamed:2003uy}
N.~Arkani-Hamed, H.-C. Cheng, M.~A. Luty, and S.~Mukohyama, {\it {Ghost
  condensation and a consistent infrared modification of gravity}},  {\em JHEP}
  {\bf 0405} (2004) 074, [\href{http://xxx.lanl.gov/abs/hep-th/0312099}{{\tt
  hep-th/0312099}}].

\bibitem{ArmendarizPicon:1999rj}
C.~Armendariz-Picon, T.~Damour, and V.~F. Mukhanov, {\it {k - inflation}},
  {\em Phys.Lett.} {\bf B458} (1999) 209--218,
  [\href{http://xxx.lanl.gov/abs/hep-th/9904075}{{\tt hep-th/9904075}}].

\bibitem{Cheung:2007st}
C.~Cheung, P.~Creminelli, A.~Fitzpatrick, J.~Kaplan, and L.~Senatore, {\it {The
  Effective Field Theory of Inflation}},  {\em JHEP} {\bf 0803} (2008) 014,
  [\href{http://xxx.lanl.gov/abs/0709.0293}{{\tt arXiv:0709.0293}}].

\bibitem{Weinberg:2008hq}
S.~Weinberg, {\it {Effective Field Theory for Inflation}},  {\em Phys.Rev.}
  {\bf D77} (2008) 123541, [\href{http://xxx.lanl.gov/abs/0804.4291}{{\tt
  arXiv:0804.4291}}].

\bibitem{Jacobson:2000xp}
T.~Jacobson and D.~Mattingly, {\it {Gravity with a dynamical preferred frame}},
   {\em Phys.Rev.} {\bf D64} (2001) 024028,
  [\href{http://xxx.lanl.gov/abs/gr-qc/0007031}{{\tt gr-qc/0007031}}].

\bibitem{Cheng:2006us}
H.-C. Cheng, M.~A. Luty, S.~Mukohyama, and J.~Thaler, {\it {Spontaneous Lorentz
  breaking at high energies}},  {\em JHEP} {\bf 0605} (2006) 076,
  [\href{http://xxx.lanl.gov/abs/hep-th/0603010}{{\tt hep-th/0603010}}].

\bibitem{Blas:2011ni}
D.~Blas and S.~Sibiryakov, {\it {Horava gravity versus thermodynamics: The
  Black hole case}},  {\em Phys.Rev.} {\bf D84} (2011) 124043,
  [\href{http://xxx.lanl.gov/abs/1110.2195}{{\tt arXiv:1110.2195}}].

\bibitem{Dubovsky:2004sg}
S.~Dubovsky, {\it {Phases of massive gravity}},  {\em JHEP} {\bf 0410} (2004)
  076, [\href{http://xxx.lanl.gov/abs/hep-th/0409124}{{\tt hep-th/0409124}}].

\bibitem{Rubakov:2004eb}
V.~Rubakov, {\it {Lorentz-violating graviton masses: Getting around ghosts, low
  strong coupling scale and VDVZ discontinuity}},
  \href{http://xxx.lanl.gov/abs/hep-th/0407104}{{\tt hep-th/0407104}}.

\bibitem{Goldberger:2008vg}
W.~D. Goldberger, {\it {AdS/CFT duality for non-relativistic field theory}},
  {\em JHEP} {\bf 0903} (2009) 069,
  [\href{http://xxx.lanl.gov/abs/0806.2867}{{\tt arXiv:0806.2867}}].

\bibitem{Duval:2008jg}
C.~Duval, M.~Hassaine, and P.~Horvathy, {\it {The Geometry of Schrodinger
  symmetry in gravity background/non-relativistic CFT}},  {\em Annals Phys.}
  {\bf 324} (2009) 1158--1167, [\href{http://xxx.lanl.gov/abs/0809.3128}{{\tt
  arXiv:0809.3128}}].

\bibitem{Herzog:2008wg}
C.~P. Herzog, M.~Rangamani, and S.~F. Ross, {\it {Heating up Galilean
  holography}},  {\em JHEP} {\bf 0811} (2008) 080,
  [\href{http://xxx.lanl.gov/abs/0807.1099}{{\tt arXiv:0807.1099}}].

\bibitem{Adams:2008wt}
A.~Adams, K.~Balasubramanian, and J.~McGreevy, {\it {Hot Spacetimes for Cold
  Atoms}},  {\em JHEP} {\bf 0811} (2008) 059,
  [\href{http://xxx.lanl.gov/abs/0807.1111}{{\tt arXiv:0807.1111}}].

\bibitem{Maldacena:2008wh}
J.~Maldacena, D.~Martelli, and Y.~Tachikawa, {\it {Comments on string theory
  backgrounds with non-relativistic conformal symmetry}},  {\em JHEP} {\bf
  0810} (2008) 072, [\href{http://xxx.lanl.gov/abs/0807.1100}{{\tt
  arXiv:0807.1100}}].

\bibitem{Rangamani:2008gi}
M.~Rangamani, S.~F. Ross, D.~Son, and E.~G. Thompson, {\it {Conformal
  non-relativistic hydrodynamics from gravity}},  {\em JHEP} {\bf 0901} (2009)
  075, [\href{http://xxx.lanl.gov/abs/0811.2049}{{\tt arXiv:0811.2049}}].

\bibitem{Seiberg:1997ad}
N.~Seiberg, {\it {Why is the matrix model correct?}},  {\em Phys.Rev.Lett.}
  {\bf 79} (1997) 3577--3580,
  [\href{http://xxx.lanl.gov/abs/hep-th/9710009}{{\tt hep-th/9710009}}].

\bibitem{Sen:1997we}
A.~Sen, {\it {D0-branes on T**n and matrix theory}},  {\em
  Adv.Theor.Math.Phys.} {\bf 2} (1998) 51--59,
  [\href{http://xxx.lanl.gov/abs/hep-th/9709220}{{\tt hep-th/9709220}}].

\bibitem{Breitenlohner:1982jf}
P.~Breitenlohner and D.~Z. Freedman, {\it {Stability in Gauged Extended
  Supergravity}},  {\em Annals Phys.} {\bf 144} (1982) 249.

\bibitem{Henkel:1993sg}
M.~Henkel, {\it {Schrodinger invariance in strongly anisotropic critical
  systems}},  {\em J.Statist.Phys.} {\bf 75} (1994) 1023--1061,
  [\href{http://xxx.lanl.gov/abs/hep-th/9310081}{{\tt hep-th/9310081}}].

\bibitem{henkel}
M.~Henkel, {\it Local scale invariance and strongly anisotropic equilibrium
  critical systems},  {\em Phys.Rev.Lett.} {\bf 78} (1997) 1940.

\bibitem{Nishida:2010tm}
Y.~Nishida and D.~T. Son, {\it {Unitary Fermi gas, epsilon expansion, and
  nonrelativistic conformal field theories}},
  \href{http://xxx.lanl.gov/abs/1004.3597}{{\tt arXiv:1004.3597}}.

\bibitem{Nakayama:2009cz}
Y.~Nakayama, M.~Sakaguchi, and K.~Yoshida, {\it {Non-Relativistic M2-brane
  Gauge Theory and New Superconformal Algebra}},  {\em JHEP} {\bf 0904} (2009)
  096, [\href{http://xxx.lanl.gov/abs/0902.2204}{{\tt arXiv:0902.2204}}].

\bibitem{Lee:2009mm}
K.-M. Lee, S.~Lee, and S.~Lee, {\it {Nonrelativistic Superconformal M2-Brane
  Theory}},  {\em JHEP} {\bf 0909} (2009) 030,
  [\href{http://xxx.lanl.gov/abs/0902.3857}{{\tt arXiv:0902.3857}}].

\bibitem{Jeong:2009aa}
J.~Jeong, H.-C. Kim, S.~Lee, E.~O~Colgain, and H.~Yavartanoo, {\it {Schrodinger
  invariant solutions of M-theory with Enhanced Supersymmetry}},  {\em JHEP}
  {\bf 1003} (2010) 034, [\href{http://xxx.lanl.gov/abs/0911.5281}{{\tt
  arXiv:0911.5281}}].

\bibitem{Karch:2002sh}
A.~Karch and E.~Katz, {\it {Adding flavor to AdS / CFT}},  {\em JHEP} {\bf
  0206} (2002) 043, [\href{http://xxx.lanl.gov/abs/hep-th/0205236}{{\tt
  hep-th/0205236}}].

\bibitem{Kobayashi:2006sb}
S.~Kobayashi, D.~Mateos, S.~Matsuura, R.~C. Myers, and R.~M. Thomson, {\it
  {Holographic phase transitions at finite baryon density}},  {\em JHEP} {\bf
  0702} (2007) 016, [\href{http://xxx.lanl.gov/abs/hep-th/0611099}{{\tt
  hep-th/0611099}}].

\bibitem{Karch:2007br}
A.~Karch and A.~O'Bannon, {\it {Holographic thermodynamics at finite baryon
  density: Some exact results}},  {\em JHEP} {\bf 0711} (2007) 074,
  [\href{http://xxx.lanl.gov/abs/0709.0570}{{\tt arXiv:0709.0570}}].

\bibitem{Ammon:2012je}
M.~Ammon, M.~Kaminski, and A.~Karch, {\it {Hyperscaling-Violation on Probe
  D-Branes}},  \href{http://xxx.lanl.gov/abs/1207.1726}{{\tt arXiv:1207.1726}}.

\end{thebibliography}\endgroup

\end{document}